%% file: main.tex
\documentclass[aps,pra,twocolumn,showpacs,landscape=false]{revtex4-1}

\input{settings/all.tex}

\makeatletter
\AtBeginDocument{\let\LS@rot\@undefined}
\makeatother

\begin{document}

\title{Peripheral circuits for ideal performance of a travelling-wave parametric amplifier}
\author{Hampus Renberg Nilsson}
\email{Hampus.Renberg.Nilsson@chalmers.se}
\author{Daryoush Shiri} 
\author{Robert Rehammar} 
\author{Anita Fadavi Roudsari} 
\author{Per Delsing} 
\address{Department of Microtechnology and Nanoscience - MC2, Chalmers University of Technology,
S-412 96 G\"oteborg, Sweden.}
\date\today

\begin{abstract}

We investigate the required peripheral circuits to enable ideal performance for a high-gain travelling-wave parametric amplifier~(TWPA) based on three-wave mixing~(3WM).
By embedding the TWPA in a network of superconducting diplexers, hybrid couplers and impedance matching networks, the amplifier can deliver a high stable gain with near-quantum-limited noise performance, with suppressed gain ripples, while eliminating the reflections of the signal, the idler and the pump as well as the transmission of all unwanted tones.
We also demonstrate a configuration where the amplifier can isolate.
We call this technique \textit{Wideband Idler Filtering}~(WIF).
The theory is supported by simulations that predict over 20\,dB gain in the 4-8\,GHz band with 10\,dB isolation for a single amplifier and 30\,dB isolation for two cascaded amplifiers.
We demonstrate how the WIF-TWPAs can be used to construct controllable isolators with over 40\,dB isolation over the full 4-8\,GHz band.

\end{abstract}

\maketitle 

\section{Introduction}

To build large-scale multiqubit quantum processors, quantum-limited amplifiers \cite{Caves1982} with high gain are desired.
They are typically built with superconducting lumped-element oscillators or transmission line resonators and have demonstrated high gain with near-quantum-limited noise performance \cite{Yamamoto2008,Bergeal2010,Roch2012,Roy2016,AumentadoRev} and have become an essential part of the circuit Quantum Electrodynamics (cQED) \cite{Girvin2008} toolbox.
While Josephson parametric amplifiers (JPAs) show high gain and near-quantum-limited noise performance \cite{Zimmer1967,Kanter1971,Feldman1975,Yurke1988,Yurke1989,Olsson1988, Beltran2008,Roy2015,Simoen2015}, they have a gain-bandwidth product limit which prevents the bandwidth to be large enough for multiplexing many qubits in large-scale quantum processors.
Such a capability is provided by travelling-wave parametric amplifiers (TWPAs)~\cite{Cullen1958,Suhl1958,Tien1958} wherein the gain-bandwidth product limit is relaxed.

The basic amplification principle of the TWPA is based on the interaction between a weak signal and a strong co-propagating wave, called the pump, when propagating through a nonlinear medium. 
Under a certain phase-matching condition, this can result in a spatially exponential growth of the signal amplitude while also generating another tone, the idler.
In the quantum regime, the TWPA can also generate the quantum phenomenon of photon entanglement \cite{Aalto2021} and, under certain conditions, signal squeezing \cite{Grimsmo2017}, \ie\ phase-sensitive amplification.

There are essentially two different kinds of TWPAs:
the kinetic inductance TWPA (KITWPA) which uses the nonlinear kinetic inductance of a superconducting transmission line \cite{LeDuc2012,Pappas2014,Pappas2016,Gao2017,Erickson2017,Katz2020,GaoPRXQ2021, Duti2021}, and the Josephson TWPA (JTWPA) which uses the nonlinear inductance of Josephson junctions \cite{Siddiqi2013,Obrien2014,Macklin2015,Martinis2015,Bell2015,Planat2020,Ranadive2021,Zorin2016,Zorin2019,Sivak2019,Mukhanov2019,Aalto2021}.
The TWPAs are further distinguished by what mixing process of microwave photons they implement, three-wave mixing (3WM) or four-wave mixing (4WM) as well as what kind of dispersion they use, typically resonant phase matching (RPM) \cite{Obrien2014,Macklin2015} or periodic modulation (PM) \cite{AnitasArticle}.
In this paper we will focus on an RPM-based JTWPA implementing 3WM, as outlined in our previous work \cite{MyTWPA3WM}.
The theory is, however, applicable for other kinds of TWPAs as well.

The ideal low noise amplifier should meet several criteria:
It provides a large bandwidth with high gain and quantum-limited noise performance and only output the desired signal.
It has minimal back-action on the signal sources, especially when the signal sources are sensitive quantum systems such as qubits.
To prevent back-propagation of noise and other tones from what comes after the amplifier, isolation is often a desired property of the amplifier.
This is usually solved by using ferrite circulators \cite{Fox1955}. While circulators work well on a small scale, they are not suitable on a large scale as they are magnetic and bulky devices.
There has been previous works about non-reciprocal quantum-limited amplifiers \cite{Abdo2014,Abdo2018}, but they typically do not have a large bandwidth.
There are also other properties that may be desired, such as a high saturation power, ease of use, scalability, robustness to stray magnetic fields, \etc., which are out of the scope of this paper.

In this paper, we investigate how to build a near-quantum-limited amplifier which, in many ways, is the ideal low noise amplifier.
By embedding the amplifier in a network of diplexers and hybrid couplers, we eliminate all reflections as well as create isolation, while only outputting one signal for each signal input.

\section{Three-wave mixing: phase matching, up-conversion and impedance matching}
\label{sec:3WM_basics}

The theory in the next sections is valid for all three-wave mixing amplifiers, but all presented numbers are based on a particular design from our previous work \cite{MyTWPA3WM}.
In this section we briefly go through the basics of 3WM, the necessities of phase matching while suppressing up-conversion, as well as impedance matching.

Three-wave mixing is a three-photon process, where either one photon down-converts into two other photons, \(\omega_3 \mapsto \omega_1 + \omega_2\), or two photons up-convert into a third photon, \(\omega_1 + \omega_2 \mapsto \omega_3\).
The desired process in a 3WM amplifier is the down-conversion of the pump to a signal and an idler, \(\omega\pump \mapsto \omega\signal + \omega\idler\).
However, it has previously been shown that 3WM also supports several undesired up-conversion processes \cite{Dixon2020}, such as up-conversion of the signal, \(\omega\signal + \omega\pump \mapsto \omega\sh{}\).
These up-conversion processes deteriorate the gain as they transfer power to other frequencies depleting the pump, the signal and the idler.
This, in turn, makes the inherently small phase mismatch to have a worse impact than previously believed.
We showed in our previous work \cite{MyTWPA3WM} that due to these up-conversion processes, 3WM cannot have exponential gain without dispersion engineering, as either phase mismatch or up-conversion processes, or both, will prevent it.

Due to the similarities between the TWPA structure and a low-pass filter, the TWPA has an intrinsic cutoff frequency.
Thus, ideally no frequencies above the cutoff frequency can enter the TWPA.
Moreover, there is a large dispersion for frequencies close to the cutoff frequency.
To enable exponential gain, one has to either work with a cutoff frequency much larger than the signal frequencies to get phase matching while using dispersion engineering to suppress the up-conversion processes \cite{AnitasArticle}, or work with a cutoff frequency close to the signal frequencies to suppress the up-conversion processes while using dispersion engineering to get phase matching \cite{MyTWPA3WM}.
In this paper we will focus on the latter as it provides the highest gain per unit cell.

The TWPA in this study consists of 100 cascaded unit cells.
The unit cell is presented in \Cref{fig:SNAIL_RPM_unitcell}.
The line inductance, as well as the nonlinear inductor enabling 3WM, is provided by a Superconducting Nonlinear Asymmetric Inductive eLement (SNAIL)~\cite{Frattini2017}, biased by a magnetic dc flux such that it provides pure 3WM and no 4WM or Kerr effect.
The SNAIL consists of a superconducting loop with several Josephson junctions, in this study 2 junctions, in one arm with a critical current \(I\crit\), and a single Josephson junction in the other arm with a critical current \(\alpha I\crit\), where \(\alpha\leq1\), in this study \(\alpha=0.44\).
The ground capacitance is provided by the RPM feature consisting of a capacitor \(C\) in parallel with a capacitively coupled \(LC\)-oscillator with inductance \(L_\text{osc}\), capacitance \(C_\text{osc}\) and coupling capacitance \(C\crit\), as outlined in Refs.~\cite{Obrien2014,Macklin2015}.

The key difference between the TWPA in Ref.~\cite{Macklin2015} and the one presented here is that the cutoff frequency here is much smaller, \(\sim\)16\,GHz, and the resonance of the RPM feature 11.8\,GHz is \textit{below} the pump frequency 12\,GHz to ensure phase matching for the 3WM process.
The RPM feature is only weakly coupled to the signal line, \(C\crit=C/10\) and \(C_\text{osc}=C\), to ensure phase matching in a wide band.
The lower frequencies are achieved by using larger capacitances, \(C+C\crit = 386\)\,fF, and smaller critical currents of the junctions, \(I\crit=1\)\,\textmu{A}.

Due to the asymmetry of the unit cell, the TWPA gets an asymmetric structure with an inductive side and a capacitive side, see \Cref{fig:AsymmetricLCchain}.
To improve impedance matching, half of the capacitance can be moved from the capacitive side to the inductive side to form a symmetric \(\pi\)-structure, or half of the inductance from the inductive side to the capacitive side to form a symmetric \(T\)-structure.

With this TWPA structure, all up-conversion processes inside the gain band 4-8\,GHz are suppressed since both the lowest signal frequency plus the pump frequency is above the cutoff frequency, 4\,GHz + 12\,GHz = 16\,GHz, and the smallest idler frequency 12\,GHz - 8\,GHz = 4\,GHz plus the pump frequency is also above the cutoff frequency, see \Cref{fig:DispersionRelation}.
At the same time, the resonance below the pump frequency ensures phase matching.
This way, we get a large exponential gain \cite{MyTWPA3WM}.

The pump amplitude could be increased to a point of more than 40\,dB gain before reaching the critical currents of the Josephson junctions inside the SNAIL.
However, impedance matching issues will appear already at lower power levels and thereby be the limiting factor. 
When connecting the TWPA input and output ports to a 50\,\textOmega\ environment, reflections will occur at the ports due to the complex and frequency-dependent impedance of the TWPA.
In \Cref{fig:Gen0_Gamma} we present the reflection coefficient for 10 unit cells of both the asymmetric and symmetric \(LC\)-chains, which is very similar to the impedance of our TWPA model except close to the resonance of the RPM feature.
These reflections will in turn make the amplifier unstable.

To mitigate the issue of impedance mismatch we add impedance matching networks to the input and output ports of the TWPA.
The impedance matching networks are built as a chain of symmetric \(\pi\)-cells, but with smaller and varying values for the inductances and capacitances, chosen and tuned such that the reflections become small~\cite{TheFilterBible}.
While they do not make the TWPA perfectly impedance matched, the reflections can be made negligibly small, see the blue dot in the middle of \Cref{fig:Gen2_Gamma}.
However, when the pump is on and its amplitude turned up there is a new kind of impedance mismatch occuring, compare the different curves in \Cref{fig:Gen2_Gamma}.
As can be seen, the TWPA is impedance matched at small pump currents, but not at larger ones.
This gives rise to gain ripple, see \Cref{fig:Gen2_Gain}.
The same phenomenon also gives rise to signal reflection, which can be problematic since the reflections occur at the exact frequency of the signal source.

The reflection coefficient and the gain presented in \Cref{fig:Gen2_Gamma,fig:Gen2_Gain}, as well as all later figures of the reflection coefficients and gains, are simulated using harmonic balance in Keysight PathWave Advanced Design System~(ADS).
We built a nonlinear model for the Josephson junctions and the SNAIL \cite{Shiri2023}, similarly to the works on KITWPAs \cite{Sweetnam2022}, and simulated the transmission and reflection of the pump, signal and the idler.

To understand where this impedance mismatch comes from, it is instructive to look at the pumpistor model \cite{Kyle2013,Kyle2014}.
The model tells us that the generation of the idler gives a perturbation to the unit cell impedance which can be modelled as a complex-valued inductance, the `pumpistor'.
The imaginary part of this complex inductance is equivalent to a negative resistance, and this negative resistance of the pumpistor both explains the gain as well as the impedance mismatch.
While the works on the pumpistor model were done for the flux-pumped TWPA, the same principles apply for the current-pumped TWPA which we use here.

\onecolumngrid

\begin{figure}[H]
    \centering
    \begin{subfigure}{0.49\textwidth}
        \centering
        \begin{tikzpicture}\input{tikzpics/SNAIL_RPM_unitcell.tikz}\end{tikzpicture}
        \caption{}
        \label{fig:SNAIL_RPM_unitcell}
    \end{subfigure}
    \begin{subfigure}{0.49\textwidth}
        \centering
        \begin{tikzpicture}\input{tikzpics/AsymmetricLCchain.tikz}\end{tikzpicture}
        \caption{}
        \label{fig:AsymmetricLCchain}
    \end{subfigure}
    \begin{subfigure}{0.49\textwidth}
        \centering
        \begin{tikzpicture}\input{tikzpics/DispersionRelation.tikz}\end{tikzpicture}
        \caption{}
        \label{fig:DispersionRelation}
    \end{subfigure}
    \begin{subfigure}{0.49\textwidth}
        \centering
        \begin{tikzpicture}\input{tikzpics/ReflectionCoefficientLCchain.tikz}\end{tikzpicture}
        \caption{}
        \label{fig:Gen0_Gamma}
    \end{subfigure}
    \begin{subfigure}{0.49\textwidth}
        \centering
        \begin{tikzpicture}\input{tikzpics/ReflectionCoefficientDesign7a.tikz}\end{tikzpicture}
        \caption{}
        \label{fig:Gen2_Gamma}
    \end{subfigure}
    \begin{subfigure}{0.49\textwidth}
        \centering
        \begin{tikzpicture}\input{tikzpics/GainDesign7a.tikz}\end{tikzpicture}
        \caption{}
        \label{fig:Gen2_Gain}
    \end{subfigure}
    \caption{\textbf{The TWPA basics.}
    \Csref{fig:SNAIL_RPM_unitcell} The TWPA unit cell with line inductance \(L\), ground capacitance \(C\), RPM coupling capacitance \(C\crit\), RPM oscillator capacitance \(C_\text{osc}\) and RPM oscillator inductance \(L_\text{osc}\).
    \Csref{fig:AsymmetricLCchain} The asymmetric \(LC\)-chain.
    \Csref{fig:DispersionRelation} The wave number \(k\) times the unit cell length \(a\) as a function of frequency with a small signal frequency (yellow), the pump frequency (orange) and the up-converted signal frequency (purple) shown.
    \Csref{fig:Gen0_Gamma} The reflection coefficient of a discrete \(LC\)-chain with \(N=10\) unit cells for frequencies from 0 up to the cutoff frequency (16\,GHz) for the asymmetric (blue and orange) and symmetric (yellow and purple) chains.
    \Csref{fig:Gen2_Gamma} Simulated reflection coefficient in the band 4-8\,GHz of the TWPA model, built from 100 unit cells (see (\subref{fig:SNAIL_RPM_unitcell})), for \(I\pump=0\)\,nA (blue), \(I\pump=60\)\,nA (orange), \(I\pump=85\)\,nA (yellow) and \(I\pump=108\)\,nA (purple).
    \Csref{fig:Gen2_Gain} Simulated gain of the same TWPA model and the same pump strengths in (\subref{fig:Gen2_Gamma}).
    }
    \label{fig:TWPAbasics}
\end{figure}

\twocolumngrid

In summary, by using a TWPA made with 100 cascaded unit cells of the structure shown in \Cref{fig:SNAIL_RPM_unitcell} and adding impedance matching networks to the input and output ports, we get an amplifier with a high gain in a wide band.
This amplifier should ideally have a quantum-limited noise performance.
However, it is not the ideal low noise amplifier, as it also has a lot of reflections at high gain, it does not isolate and it outputs the signal, the idler and the pump.
In \Cref{sec:SingleWIF} we will look at different peripheral circuits to address these issues.

\onecolumngrid

\begin{figure}[H]
    \centering
    \begin{subfigure}{0.49\textwidth}
        \centering
        \begin{tikzpicture}\input{tikzpics/BalancedAmplifier.tikz}\end{tikzpicture}
        \caption{}
        \label{fig:Gen3b}
    \end{subfigure}
    \begin{subfigure}{0.49\textwidth}
        \centering
        \begin{tikzpicture}\input{tikzpics/DiplexedAmplifier.tikz}\end{tikzpicture}
        \caption{}
        \label{fig:Gen3d}
    \end{subfigure}
    \begin{subfigure}{0.49\textwidth}
        \centering
        \begin{tikzpicture}\input{tikzpics/LayoutGen4a.tikz}\end{tikzpicture}
        \caption{}
        \label{fig:Gen4a}
    \end{subfigure}
    \begin{subfigure}{0.49\textwidth}
        \centering
        \begin{tikzpicture}\input{tikzpics/LayoutGen4b.tikz}\end{tikzpicture}
        \caption{}
        \label{fig:Gen4b}
    \end{subfigure}
    \begin{subfigure}{0.49\textwidth}
        \centering
        \begin{tikzpicture}\input{tikzpics/GainDesign7a7b7d.tikz}\end{tikzpicture}
        \caption{}
        \label{fig:Gen234_Gain}
    \end{subfigure}
    \begin{subfigure}{0.49\textwidth}
        \centering
        \begin{tikzpicture}\input{tikzpics/Design7d_IdlerReflectionCoefficient.tikz}\end{tikzpicture}
        \caption{}
        \label{fig:Gen4a_IdlerReflection}
    \end{subfigure}
    \caption{\textbf{Different TWPA peripheral circuits.}
    \Csref{fig:Gen3b} A balanced amplifier setup with the incoming wave (blue), the reflected wave (red) and the transmitted wave (purple).
    \Csref{fig:Gen3d} A diplexed TWPA with the transmission of the pump (red), the signal (blue) and the idler (purple).
    \Csref{fig:Gen4a} A diplexed \& balanced TWPA.
    \Csref{fig:Gen4b} The single layered WIF-TWPA.
    \Csref{fig:Gen234_Gain} Simulated gain profiles of the TWPA with the different peripheral circuits.
    \Csref{fig:Gen4a_IdlerReflection} The idler reflection coefficient of the balanced \& diplexed TWPA at the `Signal in' port in the band 4-8\,GHz.
    }
    \label{fig:SingleLayeredWIF}
\end{figure}

\twocolumngrid

\section{The single layered WIF-TWPA}
\label{sec:SingleWIF}

One issue is that of signal reflection and gain ripples, which is usually solved by the use of ferrite circulators.
However, it is not easy to scale up such a solution since ferrite circulators are bulky and magnetic devices and space is a limited resource in the cryostats where these amplifiers are used.
Most standard circulators are also not compatible with cryogenic operation.
Here we instead propose to build a balanced TWPA \cite{Engelbrecht1965}, \ie\ two identical TWPAs in parallel connected to each other via 90 degree hybrid couplers, see \Cref{fig:Gen3b}.
This ensures that the reflections of the two TWPAs destructively interfere towards the input port, and hence that the device remains impedance matched.
Meanwhile, assuming that the 50\Ohm\ terminations, depicted in the figure as squares with small resistors inside, are well thermally anchored to the environment, they will not add any noise apart from quantum noise.

Another issue with TWPAs is the leakage of the pump, both forwards towards the next amplifier and backwards towards the signal source.
In most works, the pump has been coupled to the signal line via a directional coupler.
While this works for the purpose of coupling the pump, it does not prevent the reflected portion of the pump from travelling towards the signal source, nor does it prevent the transmitted portion of the pump from travelling with the signal towards the next amplifier.
Due to finite directivity, the directional coupler will also transmit a portion of the pump in the direction of the signal source.
Another solution for coupling the pump is adding diplexers \cite{TheFilterBible} in front of and after the TWPA, see \Cref{fig:Gen3d}.
A diplexer both couples the pump and filters out the reflection portion of the pump.
Since the pump frequency for 3WM is typically around twice that of the signal, there is plenty of playroom for the diplexers to filter the pump.
Specifically in our case, we aim to amplify the 4 to 8\,GHz band.
Hence, the pump should be around 12\,GHz, so we use superconducting diplexers with the transition frequency of 9\,GHz. 

The balanced TWPA solves the issue of signal reflection, and the diplexed TWPA ideally suppresses the pump leakage completely.
The most straightforward way to connect them is to connect the diplexers to the balanced amplifier, see \Cref{fig:Gen4a}.
Assuming identical TWPAs, ideal hybrid couplers and diplexers, we have now eliminated all signal reflection and all pump leakage.
The simulated gain for the single TWPA, the balanced TWPA (\Cref{fig:Gen3b}) and a balanced TWPA with 9\,GHz transition frequency diplexers (\Cref{fig:Gen4a}) is presented in \Cref{fig:Gen234_Gain}.

As can be seen in \Cref{fig:Gen234_Gain}, balancing the TWPA solves the issue of gain ripples, and adding the diplexers only affects the gain when approaching the diplexer transition frequency of 9\,GHz.
If one does not want to lose any gain bandwidth, one can move the diplexer transition frequency to a higher frequency, \eg\ 10\,GHz, but that would make it more challenging to sufficiently filter out the pump.

The setup presented in \Cref{fig:Gen4a} does however not resolve the issue of idler leakage.
As shown in \Cref{fig:Gen4a_IdlerReflection}, a part of the reflected idler still leaks backwards towards the signal source.
The issue comes from the fact that the idler phase for 3WM is given by \(\phi\idler = \phi\pump - \phi\signal\) \cite{Tien1958}.
Both the pump and the signal phases increase by 90\degree\ in the lower branch of the balancing scheme in \Cref{fig:Gen4a}, hence cancelling out their effects on the idler phase.
Therefore, the idler in the lower branch of \Cref{fig:Gen4a} will be in phase with the idler in the upper branch, and once it reaches the hybrid couplers, half of the reflected idler will leak through the left coupler in \Cref{fig:Gen4a} and half of the transmitted idler will leak through the right coupler in \Cref{fig:Gen4a}.

To be able to suppress the idler leakage, we must instead make a balanced amplifier with two diplexed amplifiers, see \Cref{fig:Gen4b}.
This design we refer to as the (single layered) \textit{Wideband Idler Filtering} (WIF) TWPA. 
While this new setup requires two pump inputs, similar to the Multi-Path Interferometric Josephson Directional Amplifier (MPIJDA) \cite{Abdo2018}, it also gives an added degree of freedom: the phase difference \(\Delta\phi\pump = \phi\pump[2] - \phi\pump[1]\) between the pumps.

If the pumps are in-phase with each other, \ie\ \(\Delta\phi\pump=0\), all of the transmitted idler goes into the right 50\,\textOmega\ termination in \Cref{fig:Gen4b}.
By swapping the `signal out' with the right termination and reading out at the idler frequency, we get an isolating amplifier since any wave entering at the output port would end up in the left termination.
However, this solution does not suppress the idler reflections.

If the pumps are out-of-phase with each other, \ie\ \(\Delta\phi\pump=\pi\), all of the reflected idler goes into the left 50\,\textOmega\ termination in \Cref{fig:Gen4b}.
However, this solution does not separate the transmitted idler from the transmitted signal, and hence cannot isolate the input from the output.

To solve all these issues simultaneously, we need to use a more complicated setup, which we will outline in \Cref{sec:DoubleWIF}.

\section{The double layered WIF-TWPA}
\label{sec:DoubleWIF}

\begin{figure}[H]
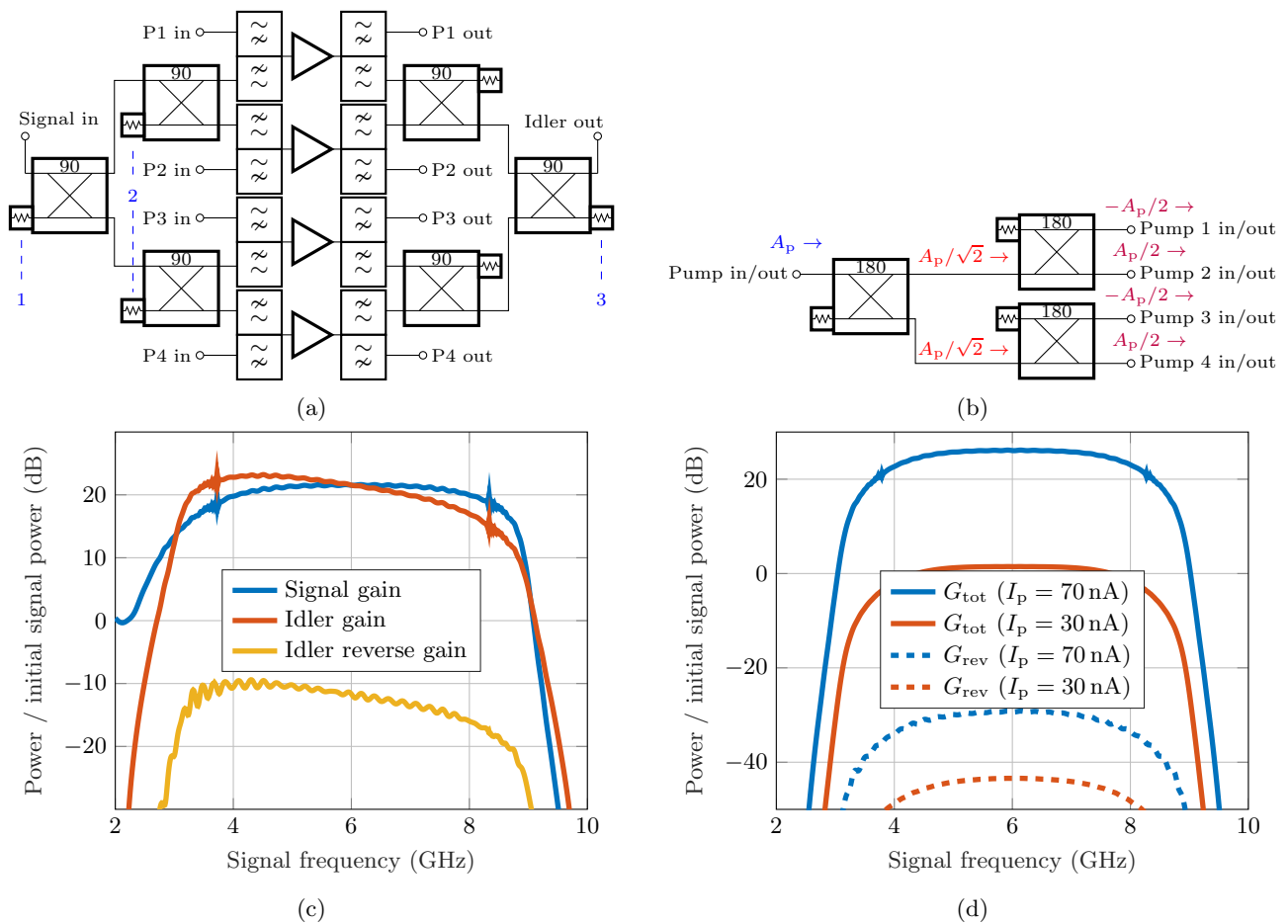

    \centering
    \begin{subfigure}{0.49\textwidth}
        \centering
        \begin{tikzpicture}\input{tikzpics/LayoutGen5a.tikz}\end{tikzpicture}
        \caption{}
        \label{fig:Gen5a}
    \end{subfigure}
    \begin{subfigure}{0.49\textwidth}
        \centering
        \begin{tikzpicture}\input{tikzpics/PumpPhaseGenerating.tikz}\end{tikzpicture}
        \caption{}
        \label{fig:PumpPhaseGen}
    \end{subfigure}
    \begin{subfigure}{0.49\textwidth}
        \centering
        \begin{tikzpicture}\input{tikzpics/Design7f_GainAndIsolation.tikz}\end{tikzpicture}
        \caption{}
        \label{fig:Design7f_GainAndIsolation}
    \end{subfigure}
    \begin{subfigure}{0.49\textwidth}
        \centering
        \begin{tikzpicture}\input{tikzpics/Design7fx2_GainAndIsolation.tikz}\end{tikzpicture}
        \caption{}
        \label{fig:Design7fx2_GainAndIsolation}
    \end{subfigure}
    \caption{\textbf{The double layered WIF-TWPA.}
    \Csref{fig:Gen5a} Schematics of the double layered WIF-TWPA.
    \Csref{fig:PumpPhaseGen} A network of couplers creating the desired pump phases of \Cref{eq:WIF_DesiredPumpPhases}.
    \Csref{fig:Design7f_GainAndIsolation} Simulated signal gain, idler gain and reverse idler gain of a double layered WIF-TWPA.
    \Csref{fig:Design7fx2_GainAndIsolation} Simulated total gain \(G_\text{tot}\) and reverse gain \(G_\text{rev}\) for two cascaded double layered WIF-TWPAs.
    }
    \label{fig:DoubleLayeredWIF}
\end{figure}

While the single layered WIF-TWPA, recall \Cref{fig:Gen4b}, suffices when the amplifier user does not need isolation or when idler reflections are tolerable, it is not good enough for all applications as it cannot filter both the reflected and the transmitted idler \textit{simultaneously}.
Here we propose a modified version that can do just that, the double layered WIF-TWPA.
It consists of two single layered WIF-TWPAs in parallel, connected to each other via another layer of hybrid couplers where we terminate the signal output, see \Cref{fig:Gen5a}.
This gives us one signal input, one idler output, four pump inputs and four pump outputs and provides three degrees of freedom:
the phases of the second, third and fourth pumps (\(\phi\pump[2],\phi\pump[3],\phi\pump[4]\)) in relation to the first (\(\phi\pump[1]\)).

Here we will focus on one combination in particular:
\begin{equation}
    \phi\pump[2] = \phi\pump[4] = \phi\pump[1]+\pi = \phi\pump[3]+\pi.
\label{eq:WIF_DesiredPumpPhases}
\end{equation}
These pump phases can be generated by using four different phase-locked signal sources and setting the phases manually.
A more convenient way, however, is to use three 180 degree hybrid couplers, similar to a Butler matrix \cite{Butler1961}, see \Cref{fig:PumpPhaseGen}, which are connected to each side of the double layered WIF-TWPA. 

With this, we get a 4-port device:
The signal input, the idler output, and only one pump input (`Pump in' in \Cref{fig:PumpPhaseGen}) and one pump output (`Pump out' in \Cref{fig:PumpPhaseGen}).
By modifying which ports are terminated, this device can be connected in several different ways giving different properties, but here we will focus on the setup presented in \Cref{fig:Gen5a} where we read out at the idler frequency.
With this setup, the pump is filtered out with the diplexers and the reflected signal and idler are caught by the terminations labeled `2' in \Cref{fig:Gen5a}.
The transmitted idler is fully separated from the transmitted signal and exits at the `Idler out' port while the signal is terminated in the termination labeled `3'.
Simultaneously, any unwanted wave entering at the output is terminated at the termination labeled `1', hence giving isolation.

If we assume ideal components, the double layered WIF-TWPA has perfect isolation when it is not pumped.
However, when we pump it, the isolation is reduced at the idler frequency due to a weak interaction between the incoming waves and the counter-propagating pump.
The isolation reduction is dependent on the pump amplitude.
When we pump it such that we get 20\,dB gain, the isolation is reduced to only about 10\,dB, see \Cref{fig:Design7f_GainAndIsolation}.
In this figure, the `idler gain' refers to the power of the idler in relation to the initial signal power, and the `idler reverse gain' refers to the power of the idler, in relation to the initial signal power, generated by a signal propagating in the direction opposite of the pump.

To improve the isolation, and also retrieve the initial signal frequency, we can cascade two of these amplifiers, see \Cref{fig:Design7fx2_GainAndIsolation}.
If we only care about isolation and do not require any gain, we can pump more weakly which improves the isolation.
Note that this setup is symmetric, \ie\ we can decide in which directions it should transmit and isolate by choosing the direction of the pump.

\onecolumngrid

\twocolumngrid

\section{Discussion and Conclusions}
\label{sec:Conclusions}

In this paper, we have discussed different issues with a high gain travelling-wave parametric amplifier based on three-wave mixing, especially the lack of isolation and the leakage of unwanted tones both in the forward and backward directions, and we have proposed solutions to these problems:
a peripheral circuit~\cite{WIFpatent} for the TWPA to create what we call ``Wideband Idler Filtering''~(WIF).
In a regular TWPA, the pump can be filtered out by using diplexers, and signal reflection by building a balanced amplifier setup, but the idler reflection and the idler transmission are mroe difficult to filter without sacrificing bandwidth.
We show that by putting the diplexers between the hybrid couplers and the two TWPAs in a balanced amplifier setup, and by setting the pump phases to appropriate values, either the reflected idler or the transmitted idler can be filtered out, but not both simultaneously.
Then we show that by using four parallel TWPAs, coupling them to each other in a suitable way and setting the pump phases to appropriate values, we can filter out both the reflected and the transmitted idler and also achieve isolation.
To further improve the isolation we can cascade two of these amplifiers.
The different peripheral circuits and their capabilities are presented in \Cref{tab:EmbeddingSummary}.

What we have not discussed in this paper is the effect of nonidealities, \ie\ what happens if the couplers, the diplexers, or the TWPAs are not performing exactly as desired.
Since we present a quite large and complicated solution with many components, there are many potential different scenarios that may occur and a study of the effects of these scenarios is out of the scope of this paper.
Here we only present a short discussion on what could happen and what we believe is the largest potential problem.

One potential problem is nonidealities in the couplers or the diplexers.
The couplers may have some phase or amplitude imbalance, and the diplexers may allow pump leakage and may couple the pumps with different amplitudes or phases, which will lead to some leakage.
How large the leakage is depends on how bad the couplers are, but with a 30~degree phase imbalance for two waves entering an ideal coupler, we only lose about 0.3\,dB due to leakage, which translates to about \(-11\)\,dB going into the wrong channel.
This is significantly improved if the imbalance is smaller.

Another potential problem is if the TWPAs are not identical, \eg\ if they do not experience the same magnetic flux density.
If one TWPA experiences a 1\,\% stronger magnetic field than the other, it shifts its cutoff frequency and hence the wave number for a given frequency, which in the centre of the gain band could lead to an increase of phase with 2~degrees per unit cell.
Since we have 100 unit cells, the accumulated phase can stretch all the way to 180~degrees, which is enough to make the couplers work fully counter-productively.
We hence conclude that nonidentical TWPAs can lead to a bigger problem.

In conclusion, while the solutions presented in this paper may seem complicated, we want to stress that this essentially makes the ideal low noise amplifier.
It has, in theory, quantum-limited noise performance, a high gain and a large bandwidth, while also both mitigating all leakage and providing isolation.
The saturation power of our device is not necessarily large enough for all applications, but since the amplifier isolates and only outputs one signal, while it does not reflect, it can be cascaded with other amplifiers to raise the saturation power.

\vspace*{5mm}
\begin{table}[H]
    \centering
    \caption{A summary of the TWPA performance using the different peripheral circuits.
    The presented isolation numbers are for the approximately lowest isolation when our TWPA model delivers at least 20\,dB gain in the full 4-8\,GHz band.}
    \label{tab:EmbeddingSummary}
    \begin{tabular}{|l|c|c|c|c|}
        \hline
         & \massatext[0.15]{Signal \mbox{reflection}} & \massatext[0.12]{Pump leakage} & \massatext[0.12]{Idler \mbox{leakage}} & Isolation \\ \hline
        \massatext[0.23]{No peripheral circuits} & Yes & Yes & Yes & - \\ \hline
        Balanced & No & Yes & Yes & - \\ \hline
        Diplexed & Yes & No & Yes & - \\ \hline
        \massatext[0.19]{Diplexed~\& balanced} & No & No & Yes & - \\ \hline
        \massatext[0.27]{Single layered WIF (\(\Delta\phi\pump=0\))} & No & No & Backwards & 10\,dB \\ \hline
        \massatext[0.275]{Single layered WIF (\(\Delta\phi\pump=\pi\))} & No & No & Forwards & - \\ \hline
        \massatext{Double layered WIF} & No & No & No & 10\,dB \\ \hline
        \massatext{Two \mbox{cascaded} double layered WIFs} & No & No & No & 30\,dB \\ \hline
        \massatext{The ideal case} & No & No & No & \(\infty\) \\ \hline
    \end{tabular}
\end{table}

\section{Acknowledgements}
The project was supported by the Knut and Alice Wallenberg foundation via the Wallenberg Centre for Quantum Technology.
The authors would like to thank Vitaly Shumeiko for useful discussions leading to the TWPA model outlined in this paper, and Christian Fager for assistance in setting up the simulation models.


\end{document}

%% file: settings/all.tex
\input{settings/packages.tex}

\input{settings/cmds.tex}

\input{settings/tikzicons.tex}
\input{settings/others.tex}

%% file: settings/packages.tex
\usepackage[dvips]{graphicx}
\usepackage{caption}
\usepackage{subcaption}

\usepackage{amsmath,amssymb}
\usepackage{amsfonts}
\usepackage{color}

\usepackage[normalem]{ulem} 
\usepackage{placeins}
\usepackage{mdframed} 
\usepackage{soul} 
\usepackage{comment} 

\usepackage{tikz}
\usepackage{pgfplots}
\usepackage{cleveref}
\usepackage{float}
\usepackage{soul}
\usepackage{lipsum}
\usepackage{readarray}
\usepackage{textgreek}
\usepgfplotslibrary{smithchart}

%% file: settings/cmds.tex
\newcommand\degree{\ensuremath{^\circ}}
\newcommand\Ohm{\,\textOmega}
\newcommand\massatext[2][0.25]{\begin{minipage}{#1\linewidth} \vspace{1mm} #2 \vspace{1mm} \end{minipage}}
\newcommand\Csref[1]{\textbf{\subref{#1}},}

\newcommand\eg{\textit{e.g.}}
\newcommand\ie{\textit{i.e.}}

\newcommand\etal{\textit{et al.}}
\newcommand\etc{\textit{etc}}

\newcounter{tikznumber}
\newcommand\I{\ensuremath{\mathrm{i}}}

\newcommand\J[1][]{\ensuremath{_{\mathrm{J}#1}}}

\newcommand\crit[1][]{\ensuremath{_{\mathrm{c}#1}}}

\newcommand\pump[1][]{\ensuremath{_{\mathrm{p}#1}}} 
\newcommand\signal[1][]{\ensuremath{_{\mathrm{s}#1}}} 
\newcommand\sh[1]{\ensuremath{_{\mathrm{s}+#1\mathrm{p}}}} 
\newcommand\sAny[1]{\ensuremath{_{\mathrm{s}#1}}} 
\newcommand\idler[1][]{\ensuremath{_{\mathrm{i}#1}}} 
\newcommand\iAny[1]{\ensuremath{_{\mathrm{i}#1}}} 

\newcommand\signed[1]{\ifnum #1>0 {+#1} \else{\ifnum #1<0 {#1} \fi} \fi}
\newcommand\coeff[1]{
    \ifnum #1=1 {+} \else{
        \ifnum #1=-1 {-} \else{
            \ifnum #1>0 {+#1} \else{
                \ifnum #1<0 {#1} \fi
            } \fi
        } \fi
    } \fi
}

%% file: settings/tikzicons.tex
\newcommand\portLR[3][IN]{\draw (#2,#3) -- (#2-0.5,#3); \draw[fill=white] (#2-0.5,#3) circle(0.05) node[anchor=east]{\scriptsize#1};}
\newcommand\portRL[3][OUT]{\draw (#2,#3) -- (#2+0.5,#3); \draw[fill=white] (#2+0.5,#3) circle(0.05) node[anchor=west]{\scriptsize#1};}
\newcommand\portUD[3][]{\draw (#2,#3) -- (#2,#3+0.5); \draw[fill=white] (#2,#3+0.5) circle(0.05) node[anchor=south]{\scriptsize#1};}
\newcommand\portDU[3][]{\draw (#2,#3) -- (#2,#3-0.5); \draw[fill=white] (#2,#3-0.5) circle(0.05) node[anchor=north]{\scriptsize#1};}

\newcommand\hybridcoupler[3][90]{\draw[very thick, fill=white] (#2-0.5,#3-0.5) rectangle(#2+0.5,#3+0.5); \foreach\y in {-0.3,0.3} {\draw (#2-0.5,#3+\y) -- (#2+0.5,#3+\y);} \draw (#2-0.3,#3+0.3) -- (#2+0.3,#3-0.3); \draw (#2-0.3,#3-0.3) -- (#2+0.3,#3+0.3); \draw (#2,#3+0.4) node{\scriptsize#1};}
\newcommand\dircoup[2]{\draw[very thick,fill=white] (#1,#2+0.3) rectangle (#1+1.5,#2-0.5); \draw (#1,#2) -- (#1+1.5,#2); \draw (#1+0.3,#2-0.5) -- (#1+0.3,#2-0.3) -- (#1+1.2,#2-0.3) -- (#1+1.2,#2-0.5); \draw (#1+1.35,#2-0.15) node{\(\Rsh\)}; \path (#1+0.75,#2) node[above=0.55]{\scriptsize Directional} node[above=0.25]{\scriptsize coupler};}
\newcommand\uddircoup[2]{\draw[very thick,fill=white] (#1,#2-0.3) rectangle (#1+1.5,#2+0.5); \draw (#1,#2) -- (#1+1.5,#2); \draw (#1+0.3,#2+0.5) -- (#1+0.3,#2+0.3) -- (#1+1.2,#2+0.3) -- (#1+1.2,#2+0.5); \draw (#1+1.35,#2+0.15) node[rotate=180]{\(\Lsh\)}; \path (#1+0.75,#2-0.3) node[below=0.05]{\scriptsize Directional} node[below=0.35]{\scriptsize coupler};}
\newcommand\realdircoup[2]{\draw[very thick,fill=white] (#1,#2+0.3) rectangle (#1+1.5,#2-0.5); \draw (#1,#2) -- (#1+1.5,#2); \draw (#1+0.3,#2-0.5) -- (#1+0.3,#2-0.3) -- (#1+1.2,#2-0.3) -- (#1+1.2,#2-0.5); \draw (#1+0.15,#2-0.07) -- (#1+0.05,#2-0.12) -- (#1+0.15,#2-0.17) (#1+0.05,#2-0.12) -- (#1+0.2,#2-0.12) -- (#1+0.2,#2-0.45) (#1+0.15,#2-0.35) -- (#1+0.2,#2-0.45) -- (#1+0.25,#2-0.35); \draw (#1+1.25,#2-0.35) -- (#1+1.3,#2-0.45) -- (#1+1.35,#2-0.35) (#1+1.3,#2-0.45) -- (#1+1.3,#2-0.12) -- (#1+1.45,#2-0.12) (#1+1.35,#2-0.17) -- (#1+1.45,#2-0.12) -- (#1+1.35,#2-0.07); \path (#1+0.75,#2) node[above=0.55]{\scriptsize Directional} node[above=0.25]{\scriptsize coupler};}
\newcommand\realuddircoup[2]{\draw[very thick,fill=white] (#1,#2-0.3) rectangle (#1+1.5,#2+0.5); \draw (#1,#2) -- (#1+1.5,#2); \draw (#1+0.3,#2+0.5) -- (#1+0.3,#2+0.3) -- (#1+1.2,#2+0.3) -- (#1+1.2,#2+0.5); \draw (#1+0.15,#2+0.07) -- (#1+0.05,#2+0.12) -- (#1+0.15,#2+0.17) (#1+0.05,#2+0.12) -- (#1+0.2,#2+0.12) -- (#1+0.2,#2+0.45) (#1+0.15,#2+0.35) -- (#1+0.2,#2+0.45) -- (#1+0.25,#2+0.35); \draw (#1+1.25,#2+0.35) -- (#1+1.3,#2+0.45) -- (#1+1.35,#2+0.35) (#1+1.3,#2+0.45) -- (#1+1.3,#2+0.12) -- (#1+1.45,#2+0.12) (#1+1.35,#2+0.17) -- (#1+1.45,#2+0.12) -- (#1+1.35,#2+0.07); \path (#1+0.75,#2-0.3) node[below=0.05]{\scriptsize Directional} node[below=0.35]{\scriptsize coupler};}
\newcommand\termuddircoup[2]{\uddircoup{#1}{#2} \vresistor{#1+0.3}{#2+1.5} \ground{#1-0.2}{#2+1.85} \draw(#1+0.1,#2+1.15) node[anchor=east]{50\,\textOmega}; \draw (#1+0.3,#2+1.5) -- (#1+0.3,#2+1.9) -- (#1-0.2,#2+1.9) -- (#1-0.2,#2+1.85) (#1+0.3,#2+0.8) -- (#1+0.3,#2+0.5);}
\newcommand\simptermuddircoup[2]{\uddircoup{#1}{#2} \miniresistorbox{#1+0.3}{#2+0.5}}
\newcommand\termdircoup[2]{\dircoup{#1}{#2} \draw (#1+0.3,#2-0.5) -- (#1+0.3,#2-0.8); \vresistor{#1+0.3}{#2-0.8} \path (#1,#2-0.8) node[anchor=north east]{50\,\textOmega}; \ground{#1+0.3}{#2-1.5}}
\newcommand\invuddircoup[2]{\uddircoup{#1}{#2}  \draw (#1+0.15,#2+0.15) node[rotate=180]{\(\Rsh\)}; \fill[white] (#1+1.25,#2+0.02) rectangle(#1+1.45,#2+0.3);}
\newcommand\terminvuddircoup[2]{\invuddircoup{#1}{#2} \vresistor{#1+1.2}{#2+1.5} \ground{#1+1.8}{#2+1.85} \draw(#1+1.4,#2+1.15) node[anchor=west]{50\,\textOmega}; \draw (#1+1.2,#2+1.5) -- (#1+1.2,#2+1.9) -- (#1+1.8,#2+1.9) -- (#1+1.8,#2+1.85) (#1+1.2,#2+0.8) -- (#1+1.2,#2+0.5);}
\newcommand\simpterminvuddircoup[2]{\invuddircoup{#1}{#2} \miniresistorbox{#1+1.2}{#2+0.5}}

\newcommand\isolator[2]{\draw[very thick,fill=white] (#1,#2-0.3) rectangle (#1+1,#2+0.3); \draw[-stealth,thick] (#1+0.2,#2) -- (#1+0.8,#2); \path (#1+0.5,#2) node[above=0.25]{\scriptsize Isolator};}
\newcommand\rightisolator[2]{\draw[very thick,fill=white] (#1,#2-0.3) rectangle (#1+1,#2+0.3); \draw[-stealth,thick] (#1+0.8,#2) -- (#1+0.2,#2); \path (#1+0.5,#2) node[above=0.25]{\scriptsize Isolator};}
\newcommand\rightdisolator[2]{\draw[very thick,fill=white] (#1,#2-0.3) rectangle (#1+1,#2+0.3); \draw[-stealth,thick] (#1+0.8,#2) -- (#1+0.2,#2); \draw[very thick,fill=white] (#1+1,#2-0.3) rectangle (#1+2,#2+0.3); \draw[-stealth,thick] (#1+1.8,#2) -- (#1+1.2,#2); \path (#1+1,#2) node[above=0.25]{\scriptsize Double isolator};}
\newcommand\visolator[2]{\draw[very thick,fill=white] (#1-0.3,#2) rectangle (#1+0.3,#2-1); \draw[-stealth,thick] (#1,#2-0.2) -- (#1,#2-0.8); \path (#1+0.3,#2-0.5) node[anchor=west]{\scriptsize Isolator};}
\newcommand\disolator[2]{\draw[very thick,fill=white] (#1,#2-0.3) rectangle (#1+1,#2+0.3); \draw[-stealth,thick] (#1+0.2,#2) -- (#1+0.8,#2); \draw[very thick,fill=white] (#1+1,#2-0.3) rectangle (#1+2,#2+0.3); \draw[-stealth,thick] (#1+1.2,#2) -- (#1+1.8,#2); \path (#1+1,#2) node[above=0.25]{\scriptsize Double isolator};}
\newcommand\circulator[2]{\draw[very thick,fill=white] (#1+0.4,#2) circle(0.4); \draw (#1+0.4,#2) node{\LARGE\(\circlearrowright\)}; \draw (#1+0.4,#2+0.4) node[anchor=south]{\scriptsize Circulator};}
\newcommand\isolcirc[2]{\draw[very thick] (#1,#2-0.3) rectangle (#1+1,#2+0.3); \draw[-stealth,thick] (#1+0.2,#2) -- (#1+0.8,#2); \draw[thick] (#1+1.4,#2) circle(0.4); \draw (#1+1.4,#2) node {\LARGE\(\circlearrowright\)}; \draw (#1+0.9,#2+0.4) node[anchor=south] {\scriptsize Isolator+Circulator};}
\newcommand\disolatordown[2]{\draw[very thick,fill=white] (#1-0.3,#2) rectangle (#1+0.3,#2-1); \draw[-stealth,thick] (#1,#2-0.2) -- (#1,#2-0.8); \draw[thick,fill=white] (#1-0.3,#2-1) rectangle (#1+0.3,#2-2); \draw[-stealth,thick] (#1,#2-1.2) -- (#1,#2-1.8); \draw (#1-0.3,#2-1) node[anchor=south east]{\scriptsize Double} node[anchor=north east]{\scriptsize isolator};}
\newcommand\disolatorup[2]{\draw[very thick,fill=white] (#1-0.3,#2) rectangle (#1+0.3,#2+1); \draw[-stealth,thick] (#1,#2+0.2) -- (#1,#2+0.8); \draw[very thick,fill=white] (#1-0.3,#2+1) rectangle (#1+0.3,#2+2); \draw[-stealth,thick] (#1,#2+1.2) -- (#1,#2+1.8); \draw (#1+0.3,#2+1) node[anchor=south west]{\scriptsize Double} node[anchor=north west]{\scriptsize isolator};}

\newcommand\amplifier[3][]{\draw[very thick,fill=white] (#2,#3-0.3) -- (#2+0.52,#3) -- (#2,#3+0.3) -- cycle; \draw (#2+0.26,#3) node[above=0.25]{\scriptsize#1};}
\newcommand\transpamplifier[3][]{\draw[very thick] (#2,#3-0.3) -- (#2+0.52,#3) -- (#2,#3+0.3) -- cycle; \draw (#2+0.26,#3) node[above=0.25]{\scriptsize#1};}
\newcommand\vamplifier[3][]{\draw[very thick,fill=white] (#2+0.3,#3) -- (#2,#3+0.52) -- (#2-0.3,#3) -- cycle; \draw (#2+0.3,#3+0.26) node[anchor=west]{\scriptsize#1};}
\newcommand\lramplifier[3][]{\draw[very thick,fill=white] (#2-0.52,#3) -- (#2,#3-0.3) -- (#2,#3+0.3) -- cycle; \draw (#2-0.26,#3) node[above=0.25]{\scriptsize#1};}
\newcommand\TWPA[2]{\amplifier[TWPA]{#1}{#2}}
\newcommand\JTWPA[2]{\TWPA{#1}{#2}}
\newcommand\HEMT[2]{\vamplifier[HEMT]{#1}{#2}}
\newcommand\hHEMT[2]{\amplifier[HEMT]{#1}{#2}}
\newcommand\lrroomamp[2]{\lramplifier{#1}{#2} \draw (#1-0.26,#2) node[above=0.5]{\scriptsize Room temp.} node[above=0.25]{\scriptsize amplifier};}
\newcommand\roomamp[2]{\vamplifier{#1}{#2} \draw (#1+0.3,#2+0.35) node[anchor=west]{\scriptsize Room temp.}; \draw (#1+0.3,#2+0.1) node[anchor=west]{\scriptsize amplifier};}
\newcommand\JTWPAmodule[2]{\draw[fill=lightgray] (#1,#2-0.7) rectangle(#1+2,#2+0.3);
\draw (#1+1,#2+0.5) node {\scriptsize JTWPA module}; \transpamplifier{#1+0.8}{#2-0.2}}
\newcommand\diplexedamplifier[2]{\draw (#1-0.2,#2) -- (#1+0.7,#2); \amplifier{#1}{#2} \diplexer{#1-0.75}{#2} \diplexer{#1+0.65}{#2}}
\newcommand\diplexedamplifierUD[2]{\draw (#1-0.2,#2) -- (#1+0.7,#2); \amplifier{#1}{#2} \diplexerUD{#1-0.75}{#2} \diplexerUD{#1+0.65}{#2}}
\newcommand\WIFTWPA[2]{\draw[very thick, fill=white] (#1-0.5,#2-0.5) rectangle(#1+0.5,#2+0.5); \amplifier{#1-0.23}{#2}}

\newcommand\notchfilter[2]{\draw[thick,fill=white] (#1,#2-0.4) rectangle (#1+0.6,#2+0.4); \draw (#1+0.3,#2) node{\(\nsim\)}; \draw (#1+0.3,#2+0.2) node{\(\sim\)}; \draw (#1+0.3,#2-0.2) node {\(\sim\)}; \draw (#1+0.3,#2+0.85) node {\scriptsize Notch}; \draw (#1+0.3,#2+0.6) node {\scriptsize filter};}
\newcommand\bandpassfilter[2]{\draw[thick,fill=white] (#1,#2-0.4) rectangle (#1+0.6,#2+0.4); \draw (#1+0.3,#2) node{\(\sim\)}; \draw (#1+0.3,#2+0.2) node{\(\nsim\)}; \draw (#1+0.3,#2-0.2) node {\(\nsim\)}; \draw (#1+0.3,#2+0.85) node {\scriptsize Band-pass}; \draw (#1+0.3,#2+0.6) node {\scriptsize filter};}
\newcommand\lowpassfilter[2]{\draw[thick,fill=white] (#1,#2-0.3) rectangle (#1+0.6,#2+0.3); \draw (#1+0.3,#2+0.1) node {\(\nsim\)}; \draw (#1+0.3,#2-0.1) node {\(\sim\)}; \draw (#1+0.3,#2+0.75) node {\scriptsize Low-pass}; \draw (#1+0.3,#2+0.5) node {\scriptsize filter};}
\newcommand\lowpassfilterWT[2]{\draw[thick,fill=white] (#1,#2-0.3) rectangle (#1+0.6,#2+0.3); \draw (#1+0.3,#2+0.1) node {\(\nsim\)}; \draw (#1+0.3,#2-0.1) node {\(\sim\)};}
\newcommand\highpassfilterWT[2]{\draw[thick,fill=white] (#1,#2-0.3) rectangle (#1+0.6,#2+0.3); \draw (#1+0.3,#2+0.1) node {\(\sim\)}; \draw (#1+0.3,#2-0.1) node {\(\nsim\)};}
\newcommand\highpassfilter[2]{\draw[thick,fill=white] (#1,#2-0.3) rectangle (#1+0.6,#2+0.3); \draw (#1+0.3,#2+0.1) node {\(\sim\)}; \draw (#1+0.3,#2-0.1) node {\(\nsim\)}; \draw (#1+0.3,#2+0.75) node {\scriptsize High-pass}; \draw (#1+0.3,#2+0.5) node {\scriptsize filter};}
\newcommand\filter[2]{\draw[thick,fill=white] (#1,#2-0.15) rectangle (#1+0.6,#2+0.15); \draw (#1+0.3,#2) node {\(\nsim\)}; \draw (#1+0.3,#2+0.35) node {\scriptsize Filter};}
\newcommand\diplexer[2]{\lowpassfilterWT{#1}{#2-0.3} \highpassfilterWT{#1}{#2+0.3}}
\newcommand\diplexerUD[2]{\lowpassfilterWT{#1}{#2+0.3} \highpassfilterWT{#1}{#2-0.3}}

\newcommand\lcryoswitch[2]{\draw[thick,fill=white] (#1,#2-0.5) rectangle (#1+0.5,#2+0.5); \draw (#1,#2) -- (#1+0.2,#2); \draw[fill] (#1+0.2,#2) circle(0.05); \draw[fill] (#1+0.3,#2+0.3) circle(0.05); \draw (#1+0.3,#2+0.3) -- (#1+0.5,#2+0.3); \draw[fill] (#1+0.3,#2-0.3) circle(0.05); \draw (#1+0.3,#2-0.3) -- (#1+0.5,#2-0.3); \draw (#1+0.2,#2) -- (#1+0.3,#2+0.3); \draw[dashed] (#1+0.2,#2) -- (#1+0.3,#2-0.3); \draw (#1+0.4,#2) node[rotate=90] {$\lcurvearrowleft$}; \draw (#1+0.25,#2) node[above=0.7]{\scriptsize Cryogenic} node[above=0.5]{\scriptsize switch};}
\newcommand\rcryoswitch[2]{\draw[thick,fill=white] (#1,#2-0.5) rectangle(#1+0.5,#2+0.5); \draw (#1,#2+0.3) -- (#1+0.2,#2+0.3); \draw[fill] (#1+0.2,#2+0.3) circle(0.05); \draw (#1,#2-0.3) -- (#1+0.2,#2-0.3); \draw[fill] (#1+0.2,#2-0.3) circle(0.05); \draw[fill] (#1+0.3,#2) circle(0.05); \draw (#1+0.3,#2) -- (#1+0.5,#2); \draw (#1+0.2,#2+0.3) -- (#1+0.3,#2); \draw[dashed] (#1+0.2,#2-0.3) -- (#1+0.3,#2); \draw (#1+0.13,#2) node[rotate=90] {$\rcurvearrowleft$};}
\newcommand\cryoswitch[2]{\lcryoswitch{#1}{#2} \rcryoswitch{#1+2.5}{#2}}
\newcommand\multiswitch[2]{\draw[fill=white] (#1,#2-0.5) rectangle(#1+0.5,#2+0.5); \draw[fill=black] (#1,#2) circle(0.05); \foreach\x in {-0.4,-0.2,0,0.2,0.4} {\draw[fill] (#1,#2) -- (#1+0.3,#2+\x) circle(0.05) -- (#1+0.5,#2+\x);} \draw (#1+0.25,#2) node[above=0.7] {\scriptsize Cryogenic} node[above=0.5]{\scriptsize switch};}
\newcommand\emptyswitch[2]{\draw[fill] (#1,#2) circle(0.05); \draw[fill] (#1+0.6,#2) circle(0.05); \draw[fill] (#1+0.3,#2+0.52) circle(0.05);}
\newcommand\lrswitch[2]{\emptyswitch{#1}{#2} \draw[dashed] (#1+0.6,#2) -- (#1+0.3,#2+0.52); \draw (#1,#2) -- (#1+0.6,#2); \draw (#1-0.05,#2+0.4) node[rotate=240]{\large$\lcurvearrowleft$};}
\newcommand\vemptyswitch[2]{\draw[fill] (#1,#2) circle(0.05); \draw[fill] (#1,#2-0.6) circle(0.05); \draw[fill] (#1-0.52,#2-0.3) circle(0.05);}
\newcommand\vswitch[2]{\vemptyswitch{#1}{#2} \draw (#1,#2) -- (#1,#2-0.6); \draw[dashed] (#1-0.52,#2-0.3) -- (#1,#2-0.6); \draw (#1-0.35,#2-0.05) node[rotate=30]{\large\(\rcurvearrowleft\)};}
\newcommand\miniswitch[2]{\draw (#1,#2-0.2) rectangle(#1+0.4,#2+0.2); \draw[fill] (#1+0.2,#2) circle(0.05); \draw (#1+0.2,#2+0.2) node[anchor=south]{\scriptsize Switch};}

\newcommand\resistor[2]{\fill[white] (#1,#2-0.2) rectangle(#1+0.7,#2+0.2); \draw (#1,#2) -- (#1+0.05,#2) -- (#1+0.1,#2+0.2) -- (#1+0.2,#2-0.2) -- (#1+0.3,#2+0.2) -- (#1+0.4,#2-0.2) -- (#1+0.5,#2+0.2) -- (#1+0.6,#2-0.2) -- (#1+0.65,#2) -- (#1+0.7,#2);}
\newcommand\vresistor[2]{\fill[white] (#1-0.2,#2) rectangle(#1+0.2,#2-0.7); \draw (#1,#2) -- (#1,#2-0.05) -- (#1+0.2,#2-0.1) -- (#1-0.2,#2-0.2) -- (#1+0.2,#2-0.3) -- (#1-0.2,#2-0.4) -- (#1+0.2,#2-0.5) -- (#1-0.2,#2-0.6) -- (#1,#2-0.65) -- (#1,#2-0.7);}
\newcommand\miniresistorbox[2]{\draw[very thick, fill=white] (#1-0.15,#2) rectangle(#1+0.15,#2+0.3); \draw (#1,#2) -- (#1,#2+0.05) -- (#1+0.05,#2+0.07) -- (#1-0.05,#2+0.11) -- (#1+0.05,#2+0.15) -- (#1-0.05,#2+0.19) -- (#1+0.05,#2+0.23) -- (#1,#2+0.25) -- (#1,#2+0.3);}
\newcommand\termination[2]{\draw[very thick, fill=white] (#1,#2-0.15) rectangle(#1+0.3,#2+0.15); \draw (#1,#2) -- (#1+0.05,#2) -- (#1+0.07,#2+0.05) -- (#1+0.11,#2-0.05) -- (#1+0.15,#2+0.05) -- (#1+0.19,#2-0.05) -- (#1+0.23,#2+0.05) -- (#1+0.25,#2) -- (#1+0.3,#2);}
\newcommand\capacitor[3][C]{\fill[white] (#2-0.2,#3) rectangle(#2+0.2,#3-0.1); \draw[thick] (#2-0.2,#3) -- (#2+0.2,#3); \draw[thick] (#2-0.2,#3-0.1) -- (#2+0.2,#3-0.1); \draw (#2+0.2,#3-0.05) node[anchor=west] {\scriptsize\(#1\)};}
\newcommand\hcapacitor[3][C]{\fill[white] (#2,#3-0.2) rectangle(#2+0.1,#3+0.2); \draw[thick] (#2,#3-0.2) -- (#2,#3+0.2); \draw[thick] (#2+0.1,#3-0.2) -- (#2+0.1,#3+0.2); \draw (#2+0.05,#3-0.2) node[anchor=north] {\scriptsize\(#1\)};}
\newcommand\inductor[3][L]{\fill[white] (#2-0.1,#3-0.1) rectangle(#2+0.2,#3-0.73); \draw (#2,#3) -- (#2,#3-0.1) to[out=-15,in=90] (#2+0.2,#3-0.2) to[out=-90,in=15] (#2,#3-0.3) -- (#2,#3-0.31) to[out=-15,in=90] (#2+0.2,#3-0.41) to[out=-90,in=15] (#2,#3-0.51) -- (#2,#3-0.52) to[out=-15,in=90] (#2+0.2,#3-0.62) to[out=-90,in=15] (#2,#3-0.72) -- (#2,#3-0.82); \draw (#2+0.2,#3-0.41) node[anchor=west]{\scriptsize\(#1\)};}
\newcommand\hinductor[3][L]{\fill[white] (#2+0.1,#3+0.2) rectangle(#2+0.73,#3-0.1); \draw (#2,#3) -- (#2+0.1,#3) to[out=85,in=180] (#2+0.2,#3+0.2) to[out=0,in=105] (#2+0.3,#3) -- (#2+0.31,#3) to[out=85,in=180] (#2+0.41,#3+0.2) to[out=0,in=105] (#2+0.51,#3) -- (#2+0.52,#3) to[out=85,in=180] (#2+0.62,#3+0.2) to[out=0,in=105] (#2+0.72,#3) -- (#2+0.82,#3); \draw (#2+0.41,#3) node[anchor=north]{\scriptsize\(#1\)};}
\newcommand\hinvinductor[3][L]{\fill[white] (#2+0.1,#3-0.2) rectangle(#2+0.73,#3+0.1); \draw (#2,#3) -- (#2+0.1,#3) to[out=-85,in=180] (#2+0.2,#3-0.2) to[out=0,in=-105] (#2+0.3,#3) -- (#2+0.31,#3) to[out=-85,in=180] (#2+0.41,#3-0.2) to[out=0,in=-105] (#2+0.51,#3) -- (#2+0.52,#3) to[out=-85,in=180] (#2+0.62,#3-0.2) to[out=0,in=-105] (#2+0.72,#3) -- (#2+0.82,#3); \draw (#2+0.41,#3) node[anchor=south]{\(#1\)};}
\newcommand\minductor[3][L]{\draw[thick,domain=#3:#3+0.3,smooth,variable=\x] plot ({0.6-13*(\x-#3-0.15)*(\x-#3-0.15)+#2-0.3},\x-0.3); \draw[thick,domain=#3+0.3:#3+0.6,smooth,variable=\x] plot ({0.6-13*(\x-#3-0.15-0.3)*(\x-#3-0.15-0.3)+#2-0.3},\x-0.9); \draw[thick,domain=#3+0.6:#3+0.9,smooth,variable=\x] plot ({0.6-13*(\x-#3-0.15-0.6)*(\x-#3-0.15-0.6)+#2-0.3},\x-1.5); \draw (#2+0.3,#3-0.45) node[anchor=west] {\scriptsize $#1$};}
\newcommand\pumpistor[2]{\draw[scale=1,thick,->,domain=0.1:0.9,smooth,variable=\x] plot({0.6*cos(360*\x)-0.15+#1},{0.6*sin(360*\x)-0.45+#2}); \inductor[L_\text{P}]{#1}{#2}}
\newcommand\impedance[3][Z]{\draw[fill=white] (#2,#3-0.1) rectangle(#2+0.5,#3+0.1); \draw (#2+0.25,#3+0.1) node[anchor=south] {\scriptsize$#1$};}
\newcommand\vimpedance[3][Z]{\draw[fill=white] (#2-0.1,#3) rectangle(#2+0.1,#3-0.5); \draw (#2+0.1,#3-0.25) node[anchor=west] {\scriptsize$#1$};}
\newcommand\biastee[2]{\draw[fill=white, very thick] (#1,#2-0.7) rectangle(#1+1.3,#2+1); \draw (#1,#2) -- (#1+1.3,#2); \draw[fill] (#1+0.9,#2) circle(0.05) -- (#1+0.9,#2+0.18); \inductor[]{#1+0.9}{#2+1} \draw (#1+0.9,#2+0.6) node[anchor=east]{\scriptsize\(L\)}; \hcapacitor{#1+0.3}{#2} \draw (#1+0.65,#2-0.7) node[anchor=north]{\scriptsize Biastee};}
\newcommand\invbiastee[2]{\draw[fill=white, very thick] (#1,#2-0.7) rectangle(#1+1.3,#2+1); \draw (#1,#2) -- (#1+1.3,#2); \draw[fill] (#1+0.3,#2) circle(0.05) -- (#1+0.3,#2+0.18); \inductor{#1+0.3}{#2+1} \hcapacitor{#1+0.9}{#2} \draw (#1+0.65,#2-0.7) node[anchor=north]{\scriptsize Biastee};}
\newcommand\terminvbiastee[2]{\invbiastee{#1}{#2} \draw (#1+0.3,#2+1) -- (#1+0.3,#2+1.3) (#1+0.3,#2+2) -- (#1+0.3,#2+2.1) -- (#1-0.5,#2+2.1) -- (#1-0.5,#2+2); \vresistor{#1+0.3}{#2+2} \draw (#1+0.5,#2+1.65) node[anchor=west]{50\,\textOmega}; \ground{#1-0.5}{#2+2}}

\newcommand\hLCparallel[4]{\draw[fill=white] (#1,#2-0.4) rectangle(#1+1,#2+0.4); \hinductor[#3]{#1+0.1}{#2+0.4} \hcapacitor[#4]{#1+0.49}{#2-0.4}}
\newcommand\vLCseries[4]{\inductor[#3]{#1}{#2} \capacitor[#4]{#1}{#2-1.3} \draw (#1,#2-0.9) -- (#1,#2-1.3);}

\newcommand\JJ[2]{\draw[thick,fill=white] (#1,#2-0.2) rectangle (#1+0.4,#2+0.2); \draw[thick] (#1,#2-0.2) -- (#1+0.4,#2+0.2); \draw[thick] (#1,#2+0.2) -- (#1+0.4,#2-0.2);}
\newcommand\JJWT[2]{\JJ{#1}{#2} \draw (#1+0.2,#2-0.2) node[anchor=north] {\scriptsize\(I\crit,C\J\)};}
\newcommand\IdealJJ[2]{\draw[thick] (#1,#2-0.2) -- (#1+0.4,#2+0.2); \draw[thick] (#1,#2+0.2) -- (#1+0.4,#2-0.2); \draw (#1+0.2,#2-0.2) node[anchor=north] {\scriptsize $I\crit$};} 
\newcommand\JJwithCJ[2]{\draw[fill=white] (#1,#2-0.5) rectangle(#1+1,#2+0.5); \hcapacitor[C\J]{#1+0.45}{#2+0.5} \IdealJJ{#1+0.3}{#2-0.5}}
\newcommand\SNAIL[2]{\draw[fill=white,smooth,samples=100,domain=0:2,variable=\t] plot({#1+0.15*\t*cos(360*\t-180)},{#2+0.15*\t*sin(360*\t-180)}) -- (#1-0.15,#2); \draw[white] (#1+0.015,#2-0.015) to[out=90,in=0] (#1-0.0135,#2+0.02) to[out=180,in=90] (#1-0.042,#2-0.015);}

\newcommand\ground[2]{\draw (#1,#2) -- (#1,#2-0.2); \draw (#1-0.3,#2-0.2) -- (#1+0.3,#2-0.2); \draw (#1-0.2,#2-0.3) -- (#1+0.2,#2-0.3); \draw (#1-0.1,#2-0.4) -- (#1+0.1,#2-0.4);}
\newcommand\generator[2]{\draw[fill=white] (#1,#2) circle(0.3) node {\(\sim\)};}
\newcommand\divider[2]{\draw (#1,#2) -- (#1+0.25,#2); \draw (#1+0.25,#2) -- (#1+0.5,#2+0.15) -- (#1+0.75,#2+0.15); \draw (#1+0.25,#2) -- (#1+0.5,#2-0.15) -- (#1+0.75,#2-0.15); \draw[thick] (#1,#2-0.25) rectangle (#1+0.75,#2+0.25); \draw (#1+0.375,#2+0.75) node{\scriptsize Power}; \draw (#1+0.375,#2+0.5) node{\scriptsize divider};}
\newcommand\combiner[2]{\draw[thick,fill=white] (#1,#2-0.25) rectangle (#1+0.6,#2+0.25); \draw (#1+0.3,#2) node{\(\Sigma\)};}
\newcommand\phaseshifter[2]{\draw[thick] (#1,#2-0.25) rectangle (#1+0.5,#2+0.25); \draw (#1+0.2,#2) node{\(\phi\)}; \draw[-stealth] (#1+0.3,#2-0.4) -- (#1+0.5,#2+0.45); \draw (#1+0.25,#2+0.85) node{\scriptsize Phase}; \draw (#1+0.25,#2+0.6) node{\scriptsize shifter};}
\newcommand\ampshifter[2]{\draw[thick] (#1,#2-0.25) rectangle (#1+0.5,#2+0.25); \draw (#1+0.2,#2) node{$A$}; \draw[-stealth] (#1+0.3,#2-0.4) -- (#1+0.5,#2+0.45); \draw (#1+0.25,#2+0.85) node{\scriptsize Amplitude}; \draw (#1+0.25,#2+0.6) node{\scriptsize shifter};}
\newcommand\mixer[2]{\draw[thick,fill=white] (#1+0.4,#2) circle(0.4); \draw (#1+0.12,#2-0.28) -- (#1+0.68,#2+0.28); \draw (#1+0.12,#2+0.28) -- (#1+0.68,#2-0.28);}
\newcommand\hIQmixer[2]{\mixer{#1}{#2} \draw (#1+0.4,#2-0.4) -- (#1+0.4,#2-0.8) node[anchor=north]{\scriptsize $Q$}; \draw (#1+0.4,#2+0.4) -- (#1+0.4,#2+0.8) node[anchor=south]{\scriptsize $I$}; \draw (#1+0.25,#2+0.25) node[anchor=south east] {\scriptsize IQ mixer};}
\newcommand\hTLin[2]{\draw[thick,fill=white] (#1,#2-0.15) rectangle(#1+0.7,#2+0.15);}
\newcommand\vTLin[2]{\draw[thick,fill=white] (#1-0.15,#2-0.7) rectangle(#1+0.15,#2);}
\newcommand\TLinText[5]{\draw (#1,#2+0.4) node[anchor=west]{\scriptsize\(Z=#3\)}; \draw (#1,#2) node[anchor=west]{\scriptsize\(E=#4\)}; \draw (#1,#2-0.4) node[anchor=west]{\scriptsize\(f=#5\)};}
\newcommand\attenuator[3][-20]{\draw[thick,fill=white] (#2-0.1,#3-0.4) rectangle(#2+0.1,#3); \draw (#2+0.1,#3-0.2) node[anchor=west] {\scriptsize $#1$\,dB};}
\newcommand\hattenuator[3][-20]{\draw[thick,fill=white] (#2,#3-0.1) rectangle(#2+0.4,#3+0.1); \draw (#2+0.2,#3+0.1) node[anchor=south]{\scriptsize \(#1\dB\)};}
\newcommand\VNA[2]{\draw[thick,fill=white] (#1,#2) rectangle (#1+2.5,#2+1.5); \draw[thick] (#1+0.2,#2+0.4) rectangle (#1+1.5,#2+1.3); \draw (#1+0.2,#2+0.6) -- (#1+0.4,#2+0.8) -- (#1+0.6,#2+0.5) -- (#1+0.8,#2+1) -- (#1+0.85,#2+0.4) (#1+1,#2+0.4) -- (#1+1.1,#2+0.7) -- (#1+1.5,#2+0.9); \foreach\x in {1.8,2.2} {\foreach\y in {0.5,0.8,1.1} {\draw[fill] (#1+\x-0.025,#2+\y) rectangle (#1+\x+0.025,#2+\y+0.05);}} \draw (#1,#2+1.5) node[anchor=south west]{\scriptsize VNA}; \draw[fill] (#1+0.7,#2+0.2) node[anchor=west]{\scriptsize1} circle(0.05); \draw[fill] (#1+1.8,#2+0.2) node[anchor=west]{\scriptsize2} circle(0.05);}
\newcommand\graybox[2]{\draw[fill=lightgray] (#1-0.2,#2-0.2) rectangle(#1+0.2,#2+0.2);}
\newcommand\digitizer[2]{\draw[fill=white] (#1,#2) rectangle(#1+1.5,#2+1); \draw (#1,#2+1) node[anchor=north west]{\scriptsize Digitizer}; \draw[fill] (#1+0.75,#2+0.3) circle(0.05);}
\newcommand\IQdigitizer[2]{\draw[fill=white] (#1,#2) rectangle(#1+1.5,#2+1); \draw (#1,#2+1) node[anchor=north west]{\scriptsize Digitizer}; \draw[fill] (#1+0.5,#2+0.3) circle(0.05); \draw[fill] (#1+1,#2+0.3) circle(0.05);}
\newcommand\AWG[2]{\draw[fill=white] (#1-0.75,#2-0.5) rectangle(#1+0.75,#2+0.5); \draw (#1,#2) node{AWG};}
\newcommand\pulsegenerator[2]{\draw[fill=white] (#1,#2) rectangle(#1+1.5,#2+1); \draw (#1+0.75,#2+1) node[anchor=south]{\scriptsize Pulse generator}; \draw[fill] (#1+0.75,#2+0.3) circle(0.05);  \draw[gray,smooth,domain=0:0.26,thin] plot({#1+0.32+\x},{0.1*sin(10*360*(\x)-200)+#2+0.6}); \draw[gray,smooth,domain=0:0.175,thin] plot({#1+1+\x},{0.19*sin(15*360*(\x)-170)+#2+0.7}); \draw (#1+0.2,#2+0.5) to[out=0,in=-100] (#1+0.3,#2+0.6) to[out=80,in=180] (#1+0.4,#2+0.7) -- (#1+0.5,#2+0.7) to[out=0,in=100] (#1+0.6,#2+0.6) to[out=-80,in=180] (#1+0.7,#2+0.5) -- (#1+0.9,#2+0.5) to[out=0,in=-100] (#1+1,#2+0.7) to[out=80,in=180] (#1+1.1,#2+0.9) to[out=0,in=100] (#1+1.2,#2+0.7) to[out=-80,in=180] (#1+1.3,#2+0.5);}
\newcommand\yokogawa[2]{\draw[fill=white] (#1,#2) rectangle(#1+1.5,#2+1); \draw[fill] (#1+1.25,#2+0.67) circle(0.05) node[anchor=south]{\scriptsize\(+\)} (#1+0.9,#2+0.67) circle(0.05); \draw[fill] (#1+1.25,#2+0.33) circle(0.05) node[anchor=north]{\scriptsize\(-\)} (#1+0.9,#2+0.33) circle(0.05); \draw (#1+0.9,#2+0.67) to[out=-30,in=210] (#1+1.25,#2+0.67); \draw (#1+0.9,#2+0.33) to[out=30,in=150] (#1+1.25,#2+0.33); \draw (#1+0.75,#2+1) node[anchor=south]{\scriptsize Yokogawa}; \draw (#1+0.1,#2+0.65) rectangle(#1+0.75,#2+0.9); \draw (#1+0.8,#2+0.78) node[anchor=east]{\tiny +0\,A};
\foreach\x in {0.1,0.28,0.47,0.65} {\foreach\y in {0.25,0.4} {\draw[fill] (#1+\x,#2+\y) rectangle(#1+\x+0.1,#2+\y+0.05);}}}
\newcommand\MeasDevice[3]{\draw[thick,fill=white] (#1,#2) rectangle(#1+2,#2+1.25); \shade[left color=white, right color=red] (#1+1,#2+0.1) -- (#1+1.8,#2+0.8) to[out=135,in=0] (#1+1,#2+1.12); \draw[thick] (#1+0.2,#2+0.8) -- (#1+1,#2+0.1) -- (#1+1.8,#2+0.8) to[out=135,in=45] (#1+0.2,#2+0.8) -- (#1+1,#2+0.1); \draw[fill] (#1+1,#2+0.2) circle(0.01); \draw[-stealth,domain=0:0.9,smooth,variable=\t] plot({#1+1+\t*cos(135-#3)},{#2+0.2+\t*sin(135-#3)}); \foreach\s in {0,10,20,30,40,50,60,70,80,90} {\draw[domain=0.7:0.75,smooth,variable=\t] plot({#1+1+\t*cos(135-\s)},{#2+0.2+\t*sin(135-\s)});}}
\newcommand\MeasDeviceNoise[4]{\draw[<->,blue,domain=#3-#4:#3+#4,smooth,variable=\t] plot({#1+1+0.45*cos(135-\t)},{#2+0.2+0.45*sin(135-\t)});}
\newcommand\coil[2]{\foreach\y in {0,0.05,0.1} {\draw (#1,#2+\y) ellipse(0.5 and 0.2);} \draw (#1,#2-0.2) node[anchor=north]{\scriptsize Coil};}
 
\newcommand\qubit[2]{\fill (#1+0.05,#2-0.2) rectangle(#1+0.45,#2-0.1); \fill (#1+0.05,#2+0.2) rectangle(#1+0.45,#2+0.1); \draw (#1+0.25,#2+0.15) -- (#1+0.25,#2-0.15); \draw[fill=white] (#1+0.22,#2-0.03) rectangle(#1+0.28,#2+0.03);} 
\newcommand\xmon[2]{\draw[fill] (#1-0.025,#2) rectangle(#1+0.025,#2-0.5); \draw[fill] (#1-0.25,#2-0.25-0.025) rectangle(#1+0.25,#2-0.25+0.025);}
\newcommand\resonator[2]{\draw (#1,#2-0.05) -- (#1+0.15,#2-0.05) to[out=0,in=90] (#1+0.25,#2-0.2) -- (#1+0.25,#2-0.3) to[out=270,in=0] (#1,#2-0.4) -- (#1-0.1,#2-0.4) to[out=180,in=180] (#1-0.1,#2-0.5) -- (#1+0.2,#2-0.5) to[out=0,in=0] (#1+0.2,#2-0.6) -- (#1-0.1,#2-0.6) to[out=180,in=180] (#1-0.1,#2-0.7) -- (#1+0.2,#2-0.7) to[out=0,in=0] (#1+0.2,#2-0.8) -- (#1-0.1,#2-0.8) to[out=180,in=180] (#1-0.1,#2-0.9) -- (#1,#2-0.9) to[out=0,in=90] (#1+0.05,#2-0.95) -- (#1+0.05,#2-1) (#1,#2-1) -- (#1+0.1,#2-1);} 
\newcommand\resqubit[2]{\qubit{#1}{#2-1.25} \resonator{#1+0.2}{#2}}
\newcommand\resxmon[2]{\xmon{#1+0.25}{#2-1.1} \resonator{#1+0.2}{#2}}
\newcommand\transmonchip[2]{\draw[fill=lightgray] (#1-0.2,#2-1.6) rectangle(#1+0.7,#2+0.1); \draw (#1-0.2,#2) -- (#1+0.7,#2); \resqubit{#1}{#2}}
\newcommand\xmonchip[2]{\draw[fill=lightgray] (#1-0.2,#2-1.7) rectangle(#1+0.7,#2+0.1); \draw (#1-0.2,#2) -- (#1+0.7,#2); \resxmon{#1}{#2}}
\newcommand\bigqubit[2]{\fill (#1+0.1,#2-0.4) rectangle(#1+0.9,#2-0.2); \fill (#1+0.1,#2+0.4) rectangle(#1+0.9,#2+0.2); \draw (#1+0.5,#2+0.3) -- (#1+0.5,#2-0.3); \draw[fill=white] (#1+0.44,#2-0.06) rectangle(#1+0.56,#2+0.06);} 
\newcommand\bigxmon[2]{\draw[fill] (#1-0.05,#2) rectangle(#1+0.05,#2-1); \draw[fill] (#1-0.5,#2-0.5-0.05) rectangle(#1+0.5,#2-0.5+0.05); \draw (#1,#2-1) -- (#1,#2-1.2) (#1-0.05,#2-1.15) -- (#1+0.05,#2-1.05) (#1-0.05,#2-1.05) -- (#1+0.05,#2-1.15);}
\newcommand\bigresonator[2]{\draw (#1,#2-0.05) -- (#1+0.3,#2-0.05) to[out=0,in=90] (#1+0.5,#2-0.4) -- (#1+0.5,#2-0.6) to[out=270,in=0] (#1,#2-0.8) -- (#1-0.2,#2-0.8) to[out=180,in=180] (#1-0.2,#2-1) -- (#1+0.4,#2-1) to[out=0,in=0] (#1+0.4,#2-1.2) -- (#1-0.2,#2-1.2) to[out=180,in=180] (#1-0.2,#2-1.4) -- (#1+0.4,#2-1.4) to[out=0,in=0] (#1+0.4,#2-1.6) -- (#1-0.2,#2-1.6) to[out=180,in=180] (#1-0.2,#2-1.8) -- (#1,#2-1.8) to[out=0,in=90] (#1+0.1,#2-1.9) -- (#1+0.1,#2-2) (#1,#2-2) -- (#1+0.2,#2-2);}
\newcommand\bigresqubit[2]{\bigqubit{#1}{#2-2.45} \bigresonator{#1+0.4}{#2}}
\newcommand\bigresxmon[2]{\bigxmon{#1+0.5}{#2-2.1} \bigresonator{#1+0.4}{#2}}
\newcommand\BasicBlochSphere{
    \draw (0,0) circle(2);
    \draw[fill] (0,0) circle(0.05);
    \draw[dashed] (0,0) ellipse(1.99 and 0.5);
    \draw[fill] (0,2) circle(0.05) node[anchor=south]{$\ket0$};
    \draw[fill] (0,-2) circle(0.05) node[anchor=north]{$\ket1$};
}

\newcommand\photon[3][0.2]{\draw[blue,thick,domain=0:1,smooth,variable=\x,-stealth] (#2,#3) -- plot ({#2+\x},{#3+#1*sin(3*360*\x+180)}) -- (#2+1.25,#3);}
\newcommand\photonold[3][]{\draw[blue, thick, snake=coil, line after snake=1mm,segment aspect=0, segment length=9pt, -stealth] (#2,#3)-- +(1.4,0); \draw (#2+0.7,#3+0.1) node[anchor=south]{#1};}
\newcommand\rotphoton[3][0]{\draw[blue, thick, snake=coil, line after snake=1mm,segment aspect=0, segment length=9pt, -stealth, rotate around={#1:(#2,#3)}] (#2,#3)-- +(1.4,0);}

\newcommand\state[4]{\draw[fill,color=#1] (#2,#3) circle(0.05); \draw[color=#1] (#2,#3) circle(#4);}
\newcommand\IQaxis[1][3]{\draw[->] (0,0) -- (#1,0) node[anchor=west]{\(I\)}; \draw[->] (0,0) -- (0,#1) node[anchor=south]{\(Q\)};}

%% file: settings/others.tex
\Crefname{equation}{Eq.}{Eqs.}
\Crefname{figure}{Fig.}{Figs.}
\Crefname{tabular}{Tab.}{Tabs.}
\setstcolor{red}
\pgfplotsset{compat=1.3,tick scale binop=\times}

\definecolor{mycolor1}{rgb}{0.00000,0.44700,0.74100}%
\definecolor{mycolor2}{rgb}{0.85000,0.32500,0.09800}%
\definecolor{mycolor3}{rgb}{0.92900,0.69400,0.12500}%
\definecolor{mycolor4}{rgb}{0.49400,0.18400,0.55600}%
\definecolor{mycolor5}{rgb}{0.46600,0.67400,0.18800}%
\definecolor{mycolor6}{rgb}{0.30100,0.74500,0.93300}%
\definecolor{mycolor7}{rgb}{0.63500,0.07800,0.18400}%
\definecolor{BLUE}{rgb}{0, 0, 1}

%% file: tikzpics/SNAIL_RPM_unitcell.tikz
    
\foreach\y in {1,-0.6} {\draw (0.3,\y) -- (3.8,\y);}
\SNAIL11
\draw (1,0.5) node{\(L\)};
\draw (1.7,1) -- +(0,-1.6);
\capacitor{1.7}{0.55}
\ground{2.125}{-0.6}
\draw (2.8,1) -- (2.8,-0.6);
\draw[fill=white] (2.3,-0.4) rectangle(3.3,0.6);
\capacitor[\ensuremath{C_\mathrm{c}}]{2.8}{0.85}
\capacitor[\ensuremath{C_\mathrm{osc}}]{2.3}{0.15}
\inductor[\ensuremath{L_\mathrm{osc}}]{3.3}{0.5}
\foreach\y in {-0.6,1} {
    \foreach\x in {0.3,3.8} {\draw[fill] (\x,\y) circle(0.05);}
}


%% file: tikzpics/AsymmetricLCchain.tikz
\foreach\x in {0,1.9} {
    \hinductor\x1
    \draw (\x+0.83,0) -- (\x+0.83,1) -- (\x+1.1,1);
    \draw (\x,0) -- (\x+1.1,0);
    \capacitor{\x+0.83}{0.6}
}

\foreach\x in {-1,4} {
    \draw (\x,0) -- (\x,1);
    \vresistor\x{0.85}
    \draw (\x+0.45,0.5) node{\(Z_0\)};
}

\foreach\y in {0,1} {
    \draw (1.5,\y) node{\(\cdots\)};
    \draw (-1,\y) -- (-0.5,\y);
    \draw (4,\y) -- (3.5,\y);
    \portLR[]0\y
    \portRL[]{1.1+1.9}\y
}

\draw[decorate, decoration = {brace}] (3,-0.25) -- (0,-0.25) node[midway, anchor=north]{\scriptsize\(N\)};

\draw (0,1.5) node[anchor=east, fill=mycolor1]{\scriptsize\textcolor{white}{\textbf{Inductive side} \(\rightarrow\)}};
\draw (3,1.5) node[anchor=west, fill=mycolor2]{\scriptsize\textcolor{white}{\(\leftarrow\) \textbf{Capacitive side}}};

%% file: tikzpics/ReflectionCoefficientDesign7a.tikz
\definecolor{mycolor1}{rgb}{0.00000,0.44700,0.74100}%
\definecolor{mycolor2}{rgb}{0.85000,0.32500,0.09800}%
\definecolor{mycolor3}{rgb}{0.92900,0.69400,0.12500}%
\definecolor{mycolor4}{rgb}{0.49400,0.18400,0.55600}%

\begin{smithchart}[
    xticklabels={{},{},{},{},{}}
]

	\addplot[color=mycolor1, line width=1, forget plot] table[row sep=crcr] {
		1.003 -8.227e-05 \\
		1.0028 -0.00051252 \\
		1.0026 -0.00087597 \\
		1.0022 -0.0011316 \\
		1.0017 -0.0012461 \\
		1.0012 -0.0011977 \\
		1.0007 -0.00097892 \\
		1.0003 -0.00059768 \\
		1 -7.7725e-05 \\
		0.99986 0.00054301 \\
		0.99991 0.0012151 \\
		1.0002 0.0018815 \\
		1.0006 0.0024824 \\
		1.0012 0.0029606 \\
		1.002 0.0032666 \\
		1.0028 0.0033639 \\
		1.0036 0.0032327 \\
		1.0044 0.0028731 \\
		1.0051 0.0023057 \\
		1.0055 0.0015715 \\
		1.0058 0.00072856 \\
		1.0058 -0.00015243 \\
		1.0055 -0.00099408 \\
		1.005 -0.0017194 \\
		1.0043 -0.0022588 \\
		1.0035 -0.0025571 \\
		1.0026 -0.0025788 \\
		1.0017 -0.0023121 \\
		1.0009 -0.0017704 \\
		1.0003 -0.00099184 \\
		0.99998 -3.7054e-05 \\
		0.99993 0.0010159 \\
		1.0002 0.0020774 \\
		1.0008 0.0030538 \\
		1.0016 0.0038549 \\
		1.0027 0.0044025 \\
		1.0039 0.0046371 \\
		1.0052 0.0045245 \\
		1.0064 0.0040598 \\
		1.0074 0.0032692 \\
		1.0082 0.0022093 \\
		1.0087 0.00096286 \\
		1.0088 -0.00036762 \\
		1.0085 -0.0016688 \\
		1.0079 -0.0028268 \\
		1.0069 -0.0037377 \\
		1.0057 -0.0043173 \\
		1.0044 -0.0045092 \\
		1.0031 -0.0042903 \\
		1.0018 -0.0036731 \\
		1.0009 -0.0027052 \\
		1.0002 -0.0014657 \\
		0.99997 -5.8667e-05 \\
		1.0002 0.001395 \\
		1.0008 0.0027672 \\
		1.0018 0.0039332 \\
		1.0031 0.0047824 \\
		1.0047 0.0052281 \\
		1.0063 0.0052161 \\
		1.008 0.0047306 \\
		1.0094 0.0037974 \\
		1.0105 0.0024832 \\
		1.0113 0.00089104 \\
		1.0115 -0.0008481 \\
		1.0112 -0.0025867 \\
		1.0105 -0.0041745 \\
		1.0093 -0.005472 \\
		1.0078 -0.0063644 \\
		1.0061 -0.0067716 \\
		1.0044 -0.006656 \\
		1.0028 -0.0060257 \\
		1.0014 -0.0049336 \\
		1.0004 -0.0034733 \\
		0.99994 -0.0017707 \\
		0.99998 2.5507e-05 \\
		1.0006 0.0017566 \\
		1.0017 0.0032662 \\
		1.0032 0.0044136 \\
		1.005 0.0050859 \\
		1.0069 0.0052088 \\
		1.0089 0.0047543 \\
		1.0107 0.0037453 \\
		1.0121 0.0022552 \\
		1.0131 0.00040295 \\
		1.0134 -0.0016569 \\
		1.0132 -0.0037478 \\
		1.0124 -0.0056881 \\
		1.0111 -0.0073079 \\
		1.0093 -0.0084655 \\
		1.0073 -0.0090605 \\
		1.0053 -0.0090428 \\
		1.0033 -0.0084165 \\
		1.0016 -0.0072392 \\
		1.0004 -0.0056162 \\
		0.9997 -0.0036913 \\
		0.9996 -0.0016352 \\
		1.0001 0.00036916 \\
		1.0013 0.002141 \\
		1.0029 0.003517 \\
		1.0048 0.0043649 \\
		1.007 0.0045966 \\
		1.0092 0.0041763 \\
		1.0112 0.0031264 \\
		1.0128 0.001527 \\
		1.0139 -0.00048975 \\
		1.0143 -0.002751 \\
		1.0141 -0.0050589 \\
		1.0132 -0.0072096 \\
		1.0118 -0.0090131 \\
		1.0099 -0.010311 \\
		1.0077 -0.010993 \\
		1.0054 -0.011004 \\
		1.0032 -0.010349 \\
		1.0014 -0.0090925 \\
		0.99998 -0.0073517 \\
		0.99917 -0.0052843 \\
		0.99902 -0.003076 \\
		0.99953 -0.00092508 \\
		1.0007 0.00097452 \\
		1.0023 0.0024489 \\
		1.0044 0.0033593 \\
		1.0066 0.0036153 \\
		1.0089 0.0031836 \\
		1.0109 0.0020933 \\
		1.0125 0.00043447 \\
		1.0136 -0.0016484 \\
		1.014 -0.0039698 \\
		1.0137 -0.0063201 \\
		1.0128 -0.0084872 \\
		1.0112 -0.010276 \\
		1.0092 -0.011529 \\
		1.0069 -0.012139 \\
		1.0046 -0.012059 \\
		1.0024 -0.011306 \\
		1.0006 -0.0099572 \\
		0.99921 -0.0081398 \\
		0.99846 -0.006022 \\
		0.99838 -0.0037965 \\
		0.99897 -0.0016645 \\
		1.0002 0.00018153 \\
		1.0018 0.001574 \\
		1.0039 0.0023851 \\
		1.0061 0.0025384 \\
		1.0082 0.0020173 \\
		1.0101 0.00086788 \\
		1.0115 -0.00080465 \\
		1.0124 -0.0028446 \\
		1.0126 -0.005061 \\
		1.0121 -0.007246 \\
		1.011 -0.0091963 \\
		1.0094 -0.010733 \\
		1.0074 -0.011719 \\
		1.0052 -0.012071 \\
		1.0031 -0.011766 \\
		1.0011 -0.010844 \\
		0.99948 -0.0093973 \\
		0.99839 -0.0075663 \\
		0.9979 -0.0055222 \\
		0.99803 -0.003453 \\
		0.99877 -0.0015468 \\
		1 2.4608e-05 \\
		1.0017 0.0011209 \\
		1.0036 0.0016459 \\
		1.0056 0.001557 \\
		1.0074 0.00086945 \\
		1.0089 -0.00034395 \\
		1.0099 -0.0019598 \\
		1.0104 -0.0038159 \\
		1.0103 -0.005728 \\
		1.0096 -0.0075083 \\
		1.0083 -0.0089851 \\
		1.0067 -0.01002 \\
		1.0049 -0.010519 \\
		1.003 -0.010444 \\
		1.0012 -0.0098125 \\
		0.99971 -0.008693 \\
		0.99863 -0.007199 \\
		0.99804 -0.0054763 \\
		0.99799 -0.0036882 \\
		0.99847 -0.0020013 \\
		0.99942 -0.00056929 \\
		1.0007 0.00048074 \\
		1.0023 0.0010601 \\
		1.0039 0.0011269 \\
		1.0054 0.00069052 \\
		1.0066 -0.00018973 \\
		1.0075 -0.0014103 \\
		1.0079 -0.0028346 \\
		1.0079 -0.0043073 \\
		1.0073 -0.005671 \\
		1.0064 -0.0067832 \\
		1.0051 -0.007531 \\
		1.0037 -0.0078416 \\
		1.0022 -0.0076893 \\
		1.0009 -0.0070959 \\
		0.99981 -0.0061274 \\
		0.99907 -0.0048856 \\
		0.99875 -0.0034973 \\
		0.99885 -0.0021003 \\
		0.99937 -0.00083054 \\
		1.0002 0.00019245 \\
		1.0013 0.00087682 \\
		1.0025 0.0011679 \\
		1.0037 0.0010538 \\
		1.0048 0.00056642 \\
		1.0056 -0.00022185 \\
		1.006 -0.0012065 \\
		1.0061 -0.0022625 \\
		1.0057 -0.0032589 \\
		1.0051 -0.0040736 \\
		1.0042 -0.0046068 \\
		1.0031 -0.0047926 \\
		1.0021 -0.0046051 \\
		1.0011 -0.0040607 \\
		1.0003 -0.003216 \\
		0.99988 -0.0021604 \\
		0.99978 -0.001006 \\
		1 0.00012419 \\
		1.0006 0.0011104 \\
		1.0015 0.0018489 \\
		1.0025 0.0022636 \\
		1.0036 0.0023153 \\
		1.0046 0.0020067 \\
		1.0055 0.0013831 \\
		1.006 0.00052801 \\
		1.0062 -0.00044586 \\
		1.006 -0.0014095 \\
		1.0054 -0.0022325 \\
		1.0045 -0.002799 \\
		1.0035 -0.003021 \\
		1.0024 -0.0028497 \\
		1.0013 -0.0022811 \\
		1.0005 -0.001357 \\
		1 -0.00016129 \\
		0.9999 0.0011887 \\
		1.0002 0.0025545 \\
		1.001 0.003791 \\
		1.0021 0.0047616 \\
		1.0034 0.0053544 \\
		1.0049 0.0054949 \\
		1.0064 0.0051567 \\
		1.0076 0.004367 \\
		1.0086 0.0032058 \\
		1.0091 0.0018001 \\
		1.009 0.00031028 \\
		1.0085 -0.0010866 \\
		1.0075 -0.0022175 \\
		1.0061 -0.0029329 \\
		1.0044 -0.0031254 \\
		1.0028 -0.002742 \\
		1.0013 -0.0017915 \\
		1.0001 -0.00034426 \\
		0.99947 0.0014735 \\
		0.9994 0.0034919 \\
		0.99996 0.0055131 \\
		1.0011 0.0073305 \\
		1.0029 0.0087493 \\
		1.005 0.0096072 \\
		1.0073 0.0097932 \\
		1.0096 0.0092625 \\
		1.0116 0.0080455 \\
		1.0131 0.0062496 \\
		1.014 0.004051 \\
		1.0142 0.0016784 \\
		1.0135 -0.00061118 \\
		1.0121 -0.0025607 \\
		1.0101 -0.0039432 \\
		1.0077 -0.0045871 \\
		1.0052 -0.0043967 \\
		1.0029 -0.0033622 \\
		1.0009 -0.0015615 \\
		0.99961 0.0008471 \\
		0.9991 0.00364 \\
		0.99949 0.0065484 \\
		1.0008 0.0092831 \\
		1.0029 0.011561 \\
		1.0057 0.013132 \\
		1.0089 0.013809 \\
		1.0122 0.013485 \\
		1.0153 0.012155 \\
		1.0179 0.0099209 \\
		1.0196 0.0069883 \\
		1.0204 0.0036476 \\
		1.02 0.000245 \\
		1.0186 -0.0028565 \\
		1.0162 -0.0053189 \\
		1.0131 -0.0068679 \\
		1.0097 -0.0073245 \\
		1.0063 -0.0066231 \\
		1.0033 -0.0048173 \\
		1.0009 -0.0020731 \\
		0.99958 0.0013487 \\
		0.99937 0.0051164 \\
		1.0004 0.0088578 \\
		1.0025 0.012193 \\
		1.0057 0.014767 \\
		1.0096 0.016289 \\
		1.0139 0.01656 \\
		1.0181 0.015502 \\
		1.0218 0.013174 \\
		1.0247 0.009775 \\
		1.0264 0.0056326 \\
		1.0266 0.0011671 \\
		1.0255 -0.0031533 \\
		1.0231 -0.0068679 \\
		1.0195 -0.0095788 \\
		1.0154 -0.010996 \\
		1.0109 -0.01097 \\
		1.0068 -0.0095001 \\
		1.0033 -0.0067359 \\
		1.0009 -0.002954 \\
		0.99978 0.0014688 \\
		1.0002 0.0060905 \\
		1.002 0.010442 \\
		1.0052 0.01407 \\
		1.0095 0.016575 \\
		1.0144 0.017655 \\
		1.0196 0.017142 \\
		1.0244 0.015027 \\
		1.0284 0.011476 \\
		1.031 0.0068221 \\
		1.0321 0.0015353 \\
		1.0315 -0.0038307 \\
		1.0292 -0.0087037 \\
		1.0255 -0.012564 \\
		1.0208 -0.015006 \\
		1.0156 -0.015786 \\
		1.0104 -0.014837 \\
		1.0059 -0.012277 \\
		1.0024 -0.008381 \\
		1.0004 -0.0035558 \\
		1.0001 0.0017047 \\
		1.0014 0.0068618 \\
		1.0044 0.011378 \\
		1.0088 0.014765 \\
		1.0141 0.016629 \\
		1.0199 0.01672 \\
		1.0254 0.01496 \\
		1.0303 0.011475 \\
		1.0338 0.0065874 \\
		1.0356 0.00079347 \\
		1.0354 -0.005293 \\
		1.0334 -0.011015 \\
		1.0296 -0.015756 \\
		1.0246 -0.019016 \\
		1.0188 -0.02047 \\
		1.013 -0.019998 \\
		1.0077 -0.017685 \\
		1.0034 -0.013805 \\
		1.0007 -0.0087824 \\
		0.99971 -0.0031471 \\
		1.0006 0.0025131 \\
		1.0034 0.0076041 \\
		1.0077 0.011577 \\
		1.0131 0.01398 \\
		1.0191 0.014508 \\
		1.0251 0.013043 \\
		1.0303 0.009683 \\
		1.0343 0.0047473 \\
		1.0364 -0.0012518 \\
		1.0366 -0.0076641 \\
		1.0346 -0.013783 \\
		1.0308 -0.018938 \\
		1.0256 -0.022579 \\
		1.0196 -0.024342 \\
		1.0134 -0.024083 \\
		1.0077 -0.021879 \\
		1.0031 -0.018012 \\
		1.0001 -0.012923 \\
		0.99882 -0.0071632 \\
		0.99954 -0.0013468 \\
		1.0021 0.0039086 \\
		1.0063 0.0080347 \\
		1.0117 0.010566 \\
		1.0176 0.011192 \\
		1.0235 0.009799 \\
		1.0287 0.0064953 \\
		1.0325 0.001616 \\
		1.0346 -0.0043084 \\
		1.0345 -0.010613 \\
		1.0325 -0.016585 \\
		1.0286 -0.021554 \\
		1.0234 -0.024987 \\
		1.0174 -0.026539 \\
		1.0113 -0.026095 \\
		1.0058 -0.023761 \\
		1.0014 -0.019842 \\
		0.99851 -0.014795 \\
		0.99747 -0.0091819 \\
		0.99831 -0.0036098 \\
		1.0009 0.0013213 \\
		1.005 0.0050757 \\
		1.0101 0.0072336 \\
		1.0157 0.0075401 \\
		1.021 0.0059408 \\
		1.0255 0.0026007 \\
		1.0287 -0.0021053 \\
		1.03 -0.0076314 \\
		1.0295 -0.013329 \\
		1.0271 -0.018533 \\
		1.0231 -0.02265 \\
	};

	\addplot[color=mycolor2, line width=1, forget plot] table[row sep=crcr] {
		1.0211 -0.21159 \\
		0.96223 -0.20144 \\
		0.9116 -0.1772 \\
		0.87126 -0.14267 \\
		0.84208 -0.10118 \\
		0.82429 -0.055465 \\
		0.81784 -0.0078263 \\
		0.82264 0.039686 \\
		0.83853 0.085063 \\
		0.86533 0.12613 \\
		0.90254 0.16038 \\
		0.94915 0.1849 \\
		1.0031 0.1965 \\
		1.0609 0.19209 \\
		1.1174 0.16951 \\
		1.1659 0.12868 \\
		1.1994 0.072632 \\
		1.2124 0.0077411 \\
		1.2027 -0.057405 \\
		1.1724 -0.11424 \\
		1.127 -0.15644 \\
		1.0734 -0.18097 \\
		1.0181 -0.18789 \\
		0.96616 -0.17927 \\
		0.92108 -0.15811 \\
		0.88482 -0.12754 \\
		0.85838 -0.090414 \\
		0.84218 -0.049229 \\
		0.83632 -0.0061426 \\
		0.84076 0.036874 \\
		0.85537 0.077878 \\
		0.87987 0.1148 \\
		0.91369 0.14535 \\
		0.95573 0.16693 \\
		1.004 0.17685 \\
		1.0553 0.17261 \\
		1.1052 0.1526 \\
		1.148 0.11695 \\
		1.178 0.068266 \\
		1.1907 0.011738 \\
		1.1841 -0.045665 \\
		1.1596 -0.096834 \\
		1.1213 -0.13621 \\
		1.0747 -0.16072 \\
		1.0253 -0.16983 \\
		0.9779 -0.16482 \\
		0.93572 -0.14796 \\
		0.90094 -0.12188 \\
		0.87477 -0.089119 \\
		0.85785 -0.051969 \\
		0.85044 -0.012494 \\
		0.8526 0.027405 \\
		0.86429 0.065865 \\
		0.88526 0.10094 \\
		0.91502 0.13049 \\
		0.9526 0.1522 \\
		0.99628 0.1636 \\
		1.0434 0.16241 \\
		1.09 0.14702 \\
		1.1314 0.11721 \\
		1.1623 0.074782 \\
		1.1783 0.023828 \\
		1.177 -0.029759 \\
		1.1586 -0.079522 \\
		1.1262 -0.11991 \\
		1.0844 -0.14732 \\
		1.0382 -0.16044 \\
		0.99231 -0.15981 \\
		0.95024 -0.14717 \\
		0.91445 -0.1248 \\
		0.88648 -0.095079 \\
		0.86722 -0.060259 \\
		0.85711 -0.022396 \\
		0.85637 0.016599 \\
		0.86504 0.054869 \\
		0.88301 0.090492 \\
		0.90991 0.12138 \\
		0.94498 0.14523 \\
		0.98678 0.15958 \\
		1.0329 0.16201 \\
		1.0799 0.15061 \\
		1.1231 0.12461 \\
		1.1573 0.085154 \\
		1.1776 0.035695 \\
		1.1808 -0.018234 \\
		1.1664 -0.070075 \\
		1.1367 -0.11375 \\
		1.096 -0.14496 \\
		1.0495 -0.1617 \\
		1.0022 -0.16407 \\
		0.95804 -0.1536 \\
		0.91984 -0.13254 \\
		0.88941 -0.10334 \\
		0.86779 -0.068363 \\
		0.85557 -0.029768 \\
		0.85303 0.010444 \\
		0.86029 0.050341 \\
		0.87731 0.087935 \\
		0.90384 0.12106 \\
		0.93921 0.14728 \\
		0.9821 0.16395 \\
		1.0302 0.16836 \\
		1.0799 0.15821 \\
		1.1264 0.13237 \\
		1.1638 0.091699 \\
		1.1867 0.039678 \\
		1.1913 -0.017752 \\
		1.1766 -0.073329 \\
		1.1452 -0.12023 \\
		1.1017 -0.15362 \\
		1.0521 -0.1713 \\
		1.0018 -0.17347 \\
		0.95506 -0.16189 \\
		0.91488 -0.13911 \\
		0.88311 -0.10781 \\
		0.8608 -0.070558 \\
		0.8485 -0.029632 \\
		0.84646 0.012863 \\
		0.85479 0.054907 \\
		0.87348 0.094404 \\
		0.90227 0.12904 \\
		0.94048 0.15618 \\
		0.98669 0.17288 \\
		1.0383 0.17615 \\
		1.0914 0.16346 \\
		1.1403 0.13365 \\
		1.1785 0.087939 \\
		1.2 0.030626 \\
		1.2011 -0.031275 \\
		1.1814 -0.089548 \\
		1.1443 -0.13691 \\
		1.0957 -0.16867 \\
		1.0421 -0.18319 \\
		0.98929 -0.18136 \\
		0.94147 -0.16555 \\
		0.90145 -0.13869 \\
		0.87082 -0.10374 \\
		0.85041 -0.063333 \\
		0.84062 -0.019827 \\
		0.84162 0.02464 \\
		0.85346 0.067978 \\
		0.87607 0.10797 \\
		0.90911 0.14211 \\
		0.95171 0.16752 \\
		1.0021 0.181 \\
		1.0571 0.1794 \\
		1.1119 0.16031 \\
		1.1601 0.12316 \\
		1.1946 0.070338 \\
		1.2096 0.0076342 \\
		1.2022 -0.056642 \\
		1.1738 -0.11376 \\
		1.1295 -0.15687 \\
		1.076 -0.1824 \\
		1.0202 -0.18995 \\
		0.96746 -0.1814 \\
		0.92149 -0.1597 \\
		0.88454 -0.12805 \\
		0.8578 -0.089422 \\
		0.84182 -0.046431 \\
		0.83681 -0.0013733 \\
		0.84283 0.043618 \\
		0.85987 0.086387 \\
		0.88772 0.12457 \\
		0.92583 0.15547 \\
		0.97291 0.17598 \\
		1.0265 0.18282 \\
		1.0827 0.17301 \\
		1.1357 0.14489 \\
		1.1783 0.099245 \\
		1.204 0.040259 \\
		1.2079 -0.024674 \\
		1.1894 -0.086572 \\
		1.152 -0.13731 \\
		1.102 -0.1716 \\
		1.0465 -0.18762 \\
		0.99183 -0.18637 \\
		0.94245 -0.17048 \\
		0.90129 -0.14317 \\
		0.86994 -0.10761 \\
		0.84917 -0.066604 \\
		0.83929 -0.022607 \\
		0.84037 0.022181 \\
		0.85239 0.065634 \\
		0.8752 0.10552 \\
		0.90835 0.13932 \\
		0.95088 0.16418 \\
		1.0009 0.17697 \\
		1.0551 0.17465 \\
		1.1086 0.15502 \\
		1.1551 0.1178 \\
		1.1876 0.065639 \\
		1.2008 0.0044736 \\
		1.1922 -0.057487 \\
		1.1635 -0.11185 \\
		1.1199 -0.15222 \\
		1.068 -0.17542 \\
		1.0144 -0.18131 \\
		0.9641 -0.17182 \\
		0.92069 -0.14989 \\
		0.88623 -0.11865 \\
		0.8618 -0.080995 \\
		0.84786 -0.039486 \\
		0.84457 0.0036115 \\
		0.85192 0.046177 \\
		0.86974 0.086073 \\
		0.89766 0.12098 \\
		0.9349 0.14832 \\
		0.97993 0.16524 \\
		1.0301 0.16888 \\
		1.0814 0.15692 \\
		1.1284 0.12841 \\
		1.1649 0.08476 \\
		1.1851 0.030301 \\
		1.1857 -0.028141 \\
		1.1667 -0.082747 \\
		1.1314 -0.12671 \\
		1.0857 -0.15575 \\
		1.0355 -0.16848 \\
		0.98642 -0.16585 \\
		0.94234 -0.15018 \\
		0.90586 -0.12433 \\
		0.87848 -0.091121 \\
		0.86092 -0.053157 \\
		0.85349 -0.012747 \\
		0.85624 0.027989 \\
		0.86905 0.066982 \\
		0.8916 0.10208 \\
		0.92324 0.13093 \\
		0.96275 0.15099 \\
		1.008 0.15965 \\
		1.0559 0.15459 \\
		1.1018 0.13447 \\
		1.1403 0.099716 \\
		1.166 0.053161 \\
		1.1748 8.0631e-05 \\
		1.1654 -0.052732 \\
		1.1393 -0.09857 \\
		1.101 -0.13238 \\
		1.0558 -0.15159 \\
		1.0091 -0.15598 \\
		0.96522 -0.14705 \\
		0.92726 -0.12716 \\
		0.89719 -0.098947 \\
		0.8761 -0.06496 \\
		0.86453 -0.027531 \\
		0.86269 0.011203 \\
		0.87055 0.049202 \\
		0.8879 0.084414 \\
		0.91422 0.11466 \\
		0.94859 0.13759 \\
		0.98934 0.15075 \\
		1.0339 0.15186 \\
		1.0785 0.13923 \\
		1.1185 0.11249 \\
		1.1488 0.073266 \\
		1.1648 0.025442 \\
		1.164 -0.025232 \\
		1.1466 -0.07234 \\
		1.1155 -0.1103 \\
		1.0751 -0.13546 \\
		1.0308 -0.1465 \\
		0.98708 -0.14398 \\
		0.94758 -0.12971 \\
		0.91477 -0.10607 \\
		0.89016 -0.075547 \\
		0.87461 -0.040495 \\
		0.86852 -0.0031087 \\
		0.87202 0.034542 \\
		0.88501 0.070413 \\
		0.90716 0.10237 \\
		0.93773 0.12813 \\
		0.97543 0.14528 \\
		1.0181 0.1514 \\
		1.0625 0.14452 \\
		1.1043 0.12366 \\
		1.1385 0.089525 \\
		1.1602 0.045067 \\
		1.1661 -0.0046677 \\
		1.155 -0.053403 \\
		1.1289 -0.095073 \\
		1.0917 -0.12517 \\
		1.0484 -0.14141 \\
		1.0041 -0.14366 \\
		0.9627 -0.1333 \\
		0.92726 -0.11254 \\
		0.89965 -0.083881 \\
		0.88099 -0.049764 \\
		0.87187 -0.012493 \\
		0.87255 0.025764 \\
		0.88304 0.062883 \\
		0.90312 0.096669 \\
		0.93217 0.12476 \\
		0.96906 0.14464 \\
		1.0118 0.1537 \\
		1.0572 0.14969 \\
		1.1011 0.13123 \\
		1.1381 0.098615 \\
		1.1631 0.054455 \\
		1.172 0.0037243 \\
		1.1633 -0.047055 \\
		1.1383 -0.091319 \\
		1.1011 -0.124 \\
		1.057 -0.14238 \\
		1.0113 -0.14609 \\
		0.96834 -0.13648 \\
		0.93137 -0.11584 \\
		0.90246 -0.086789 \\
		0.88283 -0.051927 \\
		0.87313 -0.013679 \\
		0.87366 0.025668 \\
		0.88443 0.063869 \\
		0.90522 0.098598 \\
		0.93541 0.12734 \\
		0.97376 0.14739 \\
		1.0181 0.15598 \\
		1.0651 0.1507 \\
		1.1099 0.13021 \\
		1.147 0.095031 \\
		1.1709 0.048277 \\
		1.1775 -0.0044432 \\
		1.1656 -0.056057 \\
		1.1374 -0.099767 \\
		1.0975 -0.13065 \\
		1.0517 -0.14639 \\
		1.0053 -0.14711 \\
		0.9627 -0.13456 \\
		0.92695 -0.1113 \\
		0.89992 -0.080107 \\
		0.88269 -0.043667 \\
		0.87579 -0.0044691 \\
		0.87941 0.035126 \\
		0.89348 0.072768 \\
		0.91763 0.10598 \\
		0.951 0.1321 \\
		0.99194 0.14828 \\
		1.0378 0.15179 \\
		1.0844 0.14054 \\
		1.1265 0.1139 \\
		1.1584 0.073543 \\
		1.1749 0.023853 \\
		1.1731 -0.028621 \\
		1.1537 -0.076658 \\
		1.1199 -0.11419 \\
		1.0773 -0.13754 \\
		1.0316 -0.1457 \\
		0.9877 -0.13972 \\
		0.94928 -0.12188 \\
		0.91874 -0.094928 \\
		0.8975 -0.061643 \\
		0.88629 -0.024628 \\
		0.88541 0.013676 \\
		0.89484 0.050903 \\
		0.91425 0.084638 \\
		0.9429 0.11233 \\
		0.97939 0.1313 \\
		1.0214 0.13891 \\
		1.0653 0.13304 \\
		1.1065 0.11273 \\
		1.1396 0.078966 \\
		1.1596 0.035195 \\
		1.1633 -0.01299 \\
		1.1502 -0.058981 \\
		1.1229 -0.096789 \\
		1.0859 -0.12236 \\
		1.0445 -0.13404 \\
		1.0035 -0.1323 \\
		0.9667 -0.11895 \\
		0.93668 -0.096454 \\
		0.91502 -0.067438 \\
		0.90259 -0.034444 \\
		0.89971 0.00013318 \\
		0.90637 0.033994 \\
		0.9222 0.064854 \\
		0.94642 0.090383 \\
		0.97766 0.10823 \\
		1.0138 0.11618 \\
		1.0516 0.11252 \\
		1.0873 0.096548 \\
		1.1164 0.069146 \\
		1.1349 0.033101 \\
		1.1399 -0.0071109 \\
		1.131 -0.046166 \\
		1.1099 -0.079101 \\
		1.0801 -0.10235 \\
		1.0459 -0.1142 \\
		1.0113 -0.11469 \\
		0.97961 -0.10509 \\
		0.95335 -0.087347 \\
		0.93406 -0.063705 \\
		0.9226 -0.036407 \\
		0.91932 -0.0076253 \\
		0.92417 0.020557 \\
		0.93672 0.046119 \\
		0.95614 0.067091 \\
		0.9811 0.08161 \\
		1.0097 0.088079 \\
		1.0393 0.085377 \\
		1.067 0.073247 \\
		1.0895 0.052574 \\
		1.1041 0.025516 \\
	};

	\addplot[color=mycolor3, line width=1, forget plot] table[row sep=crcr] {
		0.79844 -0.63324 \\
		0.6789 -0.51612 \\
		0.59784 -0.39787 \\
		0.54482 -0.28216 \\
		0.51331 -0.16966 \\
		0.49947 -0.059724 \\
		0.50156 0.048748 \\
		0.51954 0.15686 \\
		0.55504 0.2654 \\
		0.61152 0.37436 \\
		0.69465 0.48204 \\
		0.81254 0.58326 \\
		0.9754 0.6657 \\
		1.1905 0.70677 \\
		1.4519 0.66375 \\
		1.7085 0.50845 \\
		1.8883 0.193 \\
		1.8753 -0.19302 \\
		1.6917 -0.4886 \\
		1.4319 -0.64318 \\
		1.1831 -0.67598 \\
		0.98096 -0.63378 \\
		0.82897 -0.55365 \\
		0.71914 -0.45711 \\
		0.64223 -0.35487 \\
		0.59085 -0.25168 \\
		0.5602 -0.15495 \\
		0.54596 -0.053118 \\
		0.54889 0.052639 \\
		0.56715 0.15267 \\
		0.60304 0.25233 \\
		0.65942 0.3507 \\
		0.74087 0.44514 \\
		0.85345 0.52964 \\
		1.0033 0.59237 \\
		1.1928 0.61254 \\
		1.3998 0.56718 \\
		1.6103 0.41721 \\
		1.7476 0.16586 \\
		1.7455 -0.14409 \\
		1.6116 -0.39126 \\
		1.4078 -0.53802 \\
		1.1982 -0.58721 \\
		1.0255 -0.57132 \\
		0.87972 -0.513 \\
		0.76593 -0.42909 \\
		0.6906 -0.3451 \\
		0.63552 -0.25142 \\
		0.60053 -0.15669 \\
		0.58298 -0.061964 \\
		0.58234 0.036667 \\
		0.59711 0.13072 \\
		0.62889 0.22417 \\
		0.67996 0.31584 \\
		0.74904 0.39989 \\
		0.85578 0.48024 \\
		0.9893 0.53759 \\
		1.1472 0.5598 \\
		1.338 0.52255 \\
		1.5254 0.40271 \\
		1.6591 0.19641 \\
		1.6855 -0.071085 \\
		1.595 -0.3044 \\
		1.4282 -0.46248 \\
		1.2491 -0.53321 \\
		1.0726 -0.53863 \\
		0.92621 -0.49858 \\
		0.81217 -0.43224 \\
		0.72688 -0.35199 \\
		0.66585 -0.26491 \\
		0.62522 -0.17468 \\
		0.60223 -0.083099 \\
		0.59586 0.012785 \\
		0.60529 0.10506 \\
		0.63137 0.19717 \\
		0.67606 0.28808 \\
		0.73827 0.37211 \\
		0.83562 0.45466 \\
		0.95188 0.51584 \\
		1.108 0.54837 \\
		1.2914 0.52842 \\
		1.4803 0.43169 \\
		1.6304 0.24729 \\
		1.6882 0.00069236 \\
		1.6291 -0.24507 \\
		1.4792 -0.42794 \\
		1.2916 -0.52395 \\
		1.1099 -0.54443 \\
		0.95522 -0.51318 \\
		0.8327 -0.45117 \\
		0.74003 -0.37246 \\
		0.67287 -0.28532 \\
		0.62713 -0.19413 \\
		0.5998 -0.10103 \\
		0.58923 -0.0031455 \\
		0.59516 0.091549 \\
		0.61794 0.18666 \\
		0.65647 0.27706 \\
		0.71972 0.37033 \\
		0.80946 0.45699 \\
		0.93148 0.52914 \\
		1.0902 0.57237 \\
		1.283 0.56337 \\
		1.4895 0.47312 \\
		1.6619 0.28363 \\
		1.7363 0.017515 \\
		1.6782 -0.25396 \\
		1.5152 -0.45591 \\
		1.3101 -0.55927 \\
		1.1135 -0.57835 \\
		0.94874 -0.54162 \\
		0.82017 -0.47309 \\
		0.72418 -0.38818 \\
		0.65535 -0.29551 \\
		0.60891 -0.19943 \\
		0.58144 -0.10185 \\
		0.57096 -0.0034327 \\
		0.57689 0.095753 \\
		0.60003 0.19568 \\
		0.64267 0.2959 \\
		0.70882 0.39474 \\
		0.80425 0.48808 \\
		0.93613 0.56711 \\
		1.1105 0.61492 \\
		1.3254 0.60325 \\
		1.5562 0.49576 \\
		1.7421 0.27102 \\
		1.8041 -0.036584 \\
		1.7098 -0.33267 \\
		1.5078 -0.53256 \\
		1.2771 -0.61735 \\
		1.0695 -0.61468 \\
		0.9033 -0.55962 \\
		0.77782 -0.47744 \\
		0.6866 -0.38274 \\
		0.62289 -0.28295 \\
		0.58141 -0.1814 \\
		0.5587 -0.079283 \\
		0.55293 0.023291 \\
		0.56384 0.12661 \\
		0.59273 0.23089 \\
		0.64261 0.33575 \\
		0.71851 0.43926 \\
		0.82768 0.53634 \\
		0.97892 0.61577 \\
		1.179 0.65556 \\
		1.4222 0.61918 \\
		1.6698 0.46356 \\
		1.8376 0.17717 \\
		1.8403 -0.17085 \\
		1.6768 -0.4599 \\
		1.4316 -0.61969 \\
		1.1886 -0.66037 \\
		0.98712 -0.62421 \\
		0.83356 -0.54747 \\
		0.72154 -0.45214 \\
		0.64251 -0.34953 \\
		0.58933 -0.24477 \\
		0.55706 -0.13979 \\
		0.54273 -0.034925 \\
		0.54513 0.070197 \\
		0.56468 0.1761 \\
		0.6035 0.28298 \\
		0.66569 0.39006 \\
		0.75765 0.49435 \\
		0.88814 0.58837 \\
		1.0667 0.65596 \\
		1.2972 0.66678 \\
		1.5607 0.57531 \\
		1.7914 0.34321 \\
		1.8862 -0.0032421 \\
		1.7898 -0.34839 \\
		1.5597 -0.57835 \\
		1.2985 -0.66916 \\
		1.0699 -0.6593 \\
		0.89229 -0.59343 \\
		0.76183 -0.5012 \\
		0.66924 -0.3984 \\
		0.606 -0.29241 \\
		0.56589 -0.18618 \\
		0.54493 -0.080477 \\
		0.54112 0.024919 \\
		0.55415 0.13056 \\
		0.58536 0.23685 \\
		0.63797 0.34349 \\
		0.71731 0.44855 \\
		0.83108 0.54664 \\
		0.98859 0.62569 \\
		1.1967 0.66188 \\
		1.4479 0.61579 \\
		1.6973 0.44277 \\
		1.8531 0.13736 \\
		1.831 -0.2173 \\
		1.6454 -0.49512 \\
		1.391 -0.63566 \\
		1.1498 -0.65949 \\
		0.95503 -0.6121 \\
		0.80933 -0.52885 \\
		0.70475 -0.43009 \\
		0.63235 -0.3259 \\
		0.58514 -0.22064 \\
		0.55841 -0.11584 \\
		0.54947 -0.01167 \\
		0.5574 0.092228 \\
		0.5829 0.19618 \\
		0.62838 0.3 \\
		0.6981 0.40216 \\
		0.79838 0.49842 \\
		0.9371 0.57934 \\
		1.1209 0.62631 \\
		1.3471 0.60781 \\
		1.5861 0.48394 \\
		1.7673 0.23534 \\
		1.8061 -0.089919 \\
		1.6813 -0.38328 \\
		1.4591 -0.56333 \\
		1.2234 -0.62464 \\
		1.0213 -0.6033 \\
		0.86463 -0.53591 \\
		0.74965 -0.44638 \\
		0.66852 -0.34768 \\
		0.61422 -0.246 \\
		0.58165 -0.14396 \\
		0.56765 -0.0424 \\
		0.57079 0.058629 \\
		0.59119 0.1592 \\
		0.63053 0.259 \\
		0.69209 0.35664 \\
		0.78082 0.44851 \\
		0.9029 0.52696 \\
		1.0637 0.57731 \\
		1.262 0.57513 \\
		1.478 0.48826 \\
		1.6608 0.29475 \\
		1.7392 0.017679 \\
		1.6746 -0.26349 \\
		1.5002 -0.46626 \\
		1.2867 -0.56244 \\
		1.0878 -0.57154 \\
		0.92543 -0.52573 \\
		0.80205 -0.4503 \\
		0.71259 -0.36065 \\
		0.6509 -0.26503 \\
		0.61196 -0.16742 \\
		0.59236 -0.069527 \\
		0.59028 0.028005 \\
		0.60535 0.12488 \\
		0.63863 0.22059 \\
		0.69255 0.31374 \\
		0.7709 0.40119 \\
		0.87844 0.47654 \\
		1.0193 0.52809 \\
		1.1927 0.53677 \\
		1.385 0.47718 \\
		1.5596 0.32872 \\
		1.6603 0.099882 \\
		1.6439 -0.15559 \\
		1.5187 -0.3655 \\
		1.336 -0.48845 \\
		1.1485 -0.52691 \\
		0.98559 -0.50473 \\
		0.85628 -0.44599 \\
		0.75936 -0.36758 \\
		0.69034 -0.27952 \\
		0.64471 -0.18722 \\
		0.61907 -0.093364 \\
		0.61137 0.0007657 \\
		0.62089 0.094461 \\
		0.64819 0.18698 \\
		0.6951 0.27692 \\
		0.76461 0.36144 \\
		0.86049 0.43507 \\
		0.98608 0.48807 \\
		1.1411 0.50484 \\
		1.3154 0.4644 \\
		1.4814 0.34785 \\
		1.5936 0.15596 \\
		1.6096 -0.074442 \\
		1.5231 -0.28254 \\
		1.3692 -0.42205 \\
		1.1896 -0.49633 \\
		1.0179 -0.49014 \\
		0.88943 -0.44055 \\
		0.78688 -0.36919 \\
		0.71389 -0.28589 \\
		0.66384 -0.19648 \\
		0.63436 -0.10459 \\
		0.62314 -0.01172 \\
		0.62937 0.081157 \\
		0.65349 0.17314 \\
		0.69715 0.26283 \\
		0.76317 0.34742 \\
		0.85515 0.42165 \\
		0.97632 0.47621 \\
		1.1266 0.49616 \\
		1.2966 0.46139 \\
		1.4606 0.35323 \\
		1.5754 0.17092 \\
		1.599 -0.052307 \\
		1.5223 -0.25849 \\
		1.3765 -0.4009 \\
		1.2076 -0.46703 \\
		1.0491 -0.47012 \\
		0.91656 -0.43063 \\
		0.81316 -0.3657 \\
		0.73687 -0.28674 \\
		0.68421 -0.20052 \\
		0.65215 -0.11077 \\
		0.63864 -0.019493 \\
		0.64285 0.072144 \\
		0.66513 0.16309 \\
		0.70707 0.25178 \\
		0.77137 0.33531 \\
		0.86145 0.40833 \\
		0.98025 0.46148 \\
		1.1272 0.48 \\
		1.2924 0.44436 \\
		1.4501 0.33721 \\
		1.5582 0.1594 \\
		1.5777 -0.055618 \\
		1.5014 -0.25228 \\
		1.3603 -0.38708 \\
		1.1978 -0.44897 \\
		1.0457 -0.45072 \\
		0.91833 -0.4117 \\
		0.81902 -0.34809 \\
		0.7459 -0.27076 \\
		0.69585 -0.1862 \\
		0.66608 -0.098039 \\
		0.65481 -0.0083143 \\
		0.6614 0.081686 \\
		0.68642 0.1707 \\
		0.73158 0.25681 \\
		0.79961 0.33655 \\
		0.89366 0.40372 \\
		1.0157 0.4479 \\
		1.163 0.45348 \\
		1.3221 0.40168 \\
		1.4631 0.27972 \\
		1.5449 0.09697 \\
		1.5373 -0.10672 \\
		1.4448 -0.27929 \\
		1.3025 -0.38791 \\
		1.1493 -0.42965 \\
		1.0106 -0.41922 \\
		0.8968 -0.37436 \\
		0.80944 -0.30898 \\
		0.7465 -0.2322 \\
		0.70519 -0.14946 \\
		0.68323 -0.063884 \\
		0.67934 0.022646 \\
		0.69333 0.10868 \\
		0.72607 0.19252 \\
		0.77942 0.27145 \\
		0.85583 0.34084 \\
		0.95729 0.3929 \\
		1.083 0.41574 \\
		1.2252 0.39382 \\
		1.3639 0.313 \\
		1.4664 0.17193 \\
		1.5 -0.006433 \\
		1.4531 -0.17933 \\
		1.3447 -0.30785 \\
		1.209 -0.37626 \\
		1.0752 -0.3903 \\
		0.95916 -0.3646 \\
		0.86667 -0.31337 \\
		0.79774 -0.24708 \\
		0.75046 -0.1725 \\
		0.72271 -0.093797 \\
		0.713 -0.013562 \\
		0.72069 0.066295 \\
		0.74604 0.14387 \\
		0.79005 0.21659 \\
		0.85413 0.28047 \\
		0.93928 0.32927 \\
		1.0444 0.35392 \\
		1.1632 0.34294 \\
		1.2815 0.28573 \\
		1.3758 0.17974 \\
		1.4205 0.038541 \\
		1.4019 -0.10844 \\
		1.3276 -0.22922 \\
		1.2208 -0.30492 \\
		1.1063 -0.33373 \\
		1.001 -0.32432 \\
		0.91341 -0.28803 \\
		0.84592 -0.2346 \\
		0.79821 -0.17109 \\
		0.76899 -0.1023 \\
		0.76073 -0.03363 \\
		0.76403 0.037481 \\
		0.78496 0.10438 \\
		0.82171 0.16582 \\
		0.87574 0.21932 \\
		0.94549 0.25906 \\
		1.0282 0.27534 \\
		1.1165 0.26532 \\
		1.2036 0.22383 \\
		1.2751 0.14977 \\
		1.3242 0.05001 \\
	};

	\addplot[color=mycolor4, line width=1, forget plot] table[row sep=crcr] {
		0.020615 -0.84316 \\
		0.022159 -0.61987 \\
		0.024035 -0.43363 \\
		0.026551 -0.26998 \\
		0.029897 -0.11913 \\
		0.034359 0.026155 \\
		0.040374 0.1721 \\
		0.04862 0.32502 \\
		0.060349 0.49237 \\
		0.077747 0.68465 \\
		0.10506 0.91802 \\
		0.15143 1.2206 \\
		0.2396 1.6485 \\
		0.43795 2.3243 \\
		1.0271 3.5992 \\
		4.2006 6.3797 \\
		11.296 -5.202 \\
		2.0518 -4.7218 \\
		0.73121 -2.8402 \\
		0.38159 -1.935 \\
		0.24271 -1.4072 \\
		0.17493 -1.0527 \\
		0.13794 -0.78996 \\
		0.11672 -0.58039 \\
		0.1046 -0.40301 \\
		0.098544 -0.24504 \\
		0.097134 -0.098042 \\
		0.099972 0.044382 \\
		0.10709 0.18843 \\
		0.11963 0.33958 \\
		0.13941 0.50532 \\
		0.17094 0.69538 \\
		0.22147 0.9248 \\
		0.30827 1.2177 \\
		0.47096 1.618 \\
		0.8245 2.2103 \\
		1.7643 3.106 \\
		4.7793 3.5709 \\
		6.8169 -1.9654 \\
		2.7054 -3.4289 \\
		1.1783 -2.4989 \\
		0.65008 -1.8132 \\
		0.42028 -1.3562 \\
		0.30276 -1.0308 \\
		0.23632 -0.78171 \\
		0.19627 -0.57995 \\
		0.17181 -0.40737 \\
		0.15743 -0.25333 \\
		0.15069 -0.10916 \\
		0.15031 0.030732 \\
		0.15621 0.17172 \\
		0.16938 0.31932 \\
		0.19214 0.48001 \\
		0.2291 0.66231 \\
		0.28919 0.87855 \\
		0.39083 1.1478 \\
		0.57563 1.5007 \\
		0.95021 1.9812 \\
		1.8282 2.5902 \\
		3.9625 2.6336 \\
		5.6442 -0.64219 \\
		3.155 -2.7298 \\
		1.528 -2.3468 \\
		0.86148 -1.793 \\
		0.55607 -1.3715 \\
		0.39646 -1.0566 \\
		0.30463 -0.81107 \\
		0.2484 -0.60992 \\
		0.21293 -0.43729 \\
		0.19103 -0.28207 \\
		0.17817 -0.13864 \\
		0.17322 0.00068098 \\
		0.17542 0.14049 \\
		0.1854 0.28605 \\
		0.20517 0.44348 \\
		0.23888 0.62067 \\
		0.29461 0.82884 \\
		0.38887 1.0851 \\
		0.55806 1.4161 \\
		0.89238 1.8602 \\
		1.6431 2.4264 \\
		3.4511 2.6527 \\
		5.5481 0.10152 \\
		3.5688 -2.5628 \\
		1.7349 -2.4108 \\
		0.95679 -1.8682 \\
		0.60425 -1.4329 \\
		0.42233 -1.1058 \\
		0.31843 -0.85137 \\
		0.25484 -0.64388 \\
		0.21434 -0.46665 \\
		0.18837 -0.30861 \\
		0.17281 -0.16118 \\
		0.1644 -0.020466 \\
		0.16331 0.12111 \\
		0.16948 0.26823 \\
		0.18448 0.42721 \\
		0.21182 0.60631 \\
		0.25849 0.81751 \\
		0.33902 1.0796 \\
		0.48621 1.4241 \\
		0.78436 1.9038 \\
		1.4887 2.5802 \\
		3.4254 3.131 \\
		6.3603 0.25878 \\
		3.721 -3.0475 \\
		1.6359 -2.6246 \\
		0.86384 -1.9562 \\
		0.53418 -1.4734 \\
		0.36864 -1.1258 \\
		0.27534 -0.86112 \\
		0.21852 -0.64781 \\
		0.1823 -0.46675 \\
		0.15889 -0.30579 \\
		0.14427 -0.15647 \\
		0.1364 -0.012284 \\
		0.13449 0.13249 \\
		0.13873 0.28371 \\
		0.15047 0.44835 \\
		0.1728 0.63585 \\
		0.21211 0.86048 \\
		0.28213 1.1459 \\
		0.41559 1.5352 \\
		0.70451 2.1131 \\
		1.4762 3.0363 \\
		4.1915 4.0381 \\
		7.8445 -1.4133 \\
		2.9589 -3.788 \\
		1.1887 -2.6861 \\
		0.62731 -1.9128 \\
		0.39176 -1.4185 \\
		0.27385 -1.0736 \\
		0.20684 -0.81357 \\
		0.16585 -0.60377 \\
		0.13961 -0.42498 \\
		0.12267 -0.26499 \\
		0.11222 -0.11535 \\
		0.1069 0.030576 \\
		0.10627 0.17883 \\
		0.11068 0.33585 \\
		0.12154 0.50967 \\
		0.14195 0.71179 \\
		0.17848 0.96055 \\
		0.24599 1.2885 \\
		0.38307 1.7605 \\
		0.71412 2.5238 \\
		1.8059 3.9336 \\
		7.3315 4.9333 \\
		6.1422 -5.1569 \\
		1.6483 -3.6842 \\
		0.7074 -2.4153 \\
		0.39803 -1.7131 \\
		0.26161 -1.2695 \\
		0.19014 -0.95658 \\
		0.14864 -0.71562 \\
		0.12277 -0.51894 \\
		0.10619 -0.34824 \\
		0.095777 -0.19226 \\
		0.090492 -0.044047 \\
		0.087989 0.10365 \\
		0.090009 0.25642 \\
		0.096692 0.42192 \\
		0.1094 0.60951 \\
		0.13336 0.83508 \\
		0.17534 1.1222 \\
		0.25705 1.5212 \\
		0.43867 2.1389 \\
		0.95933 3.2575 \\
		3.4088 5.6045 \\
		12.1739 -2.4036 \\
		2.5078 -4.9119 \\
		0.84265 -2.9455 \\
		0.42405 -1.9924 \\
		0.2627 -1.4427 \\
		0.18462 -1.0767 \\
		0.14147 -0.80703 \\
		0.11569 -0.59248 \\
		0.099734 -0.41094 \\
		0.090029 -0.24908 \\
		0.084813 -0.097857 \\
		0.083265 0.049628 \\
		0.085568 0.2002 \\
		0.091793 0.36009 \\
		0.10382 0.53836 \\
		0.12463 0.7475 \\
		0.1608 1.0081 \\
		0.22755 1.3578 \\
		0.36608 1.8752 \\
		0.71953 2.7528 \\
		2.0465 4.5256 \\
		10.3355 4.5305 \\
		4.3552 -5.7875 \\
		1.1708 -3.4387 \\
		0.52715 -2.2433 \\
		0.30723 -1.593 \\
		0.20816 -1.1782 \\
		0.15602 -0.88187 \\
		0.12604 -0.65163 \\
		0.10814 -0.46056 \\
		0.097706 -0.29301 \\
		0.092512 -0.13879 \\
		0.091588 0.0096422 \\
		0.094777 0.15876 \\
		0.10266 0.31514 \\
		0.11679 0.4867 \\
		0.14036 0.68437 \\
		0.17993 0.92523 \\
		0.25004 1.2391 \\
		0.38747 1.6839 \\
		0.70648 2.3874 \\
		1.6953 3.6395 \\
		6.258 4.7452 \\
		6.7643 -4.5264 \\
		1.8392 -3.7115 \\
		0.76579 -2.4342 \\
		0.42303 -1.716 \\
		0.27603 -1.2637 \\
		0.20127 -0.94571 \\
		0.15933 -0.70234 \\
		0.13473 -0.50306 \\
		0.12055 -0.33043 \\
		0.11349 -0.17336 \\
		0.11206 -0.023891 \\
		0.11591 0.12453 \\
		0.12559 0.27826 \\
		0.14305 0.44498 \\
		0.17119 0.63278 \\
		0.21801 0.85772 \\
		0.29838 1.1421 \\
		0.44874 1.5278 \\
		0.77073 2.0935 \\
		1.6192 2.9609 \\
		4.4029 3.5966 \\
		6.9629 -1.6659 \\
		2.7435 -3.4556 \\
		1.1527 -2.4974 \\
		0.62077 -1.7934 \\
		0.39465 -1.3286 \\
		0.28105 -0.99891 \\
		0.21779 -0.74705 \\
		0.18065 -0.54194 \\
		0.15885 -0.36551 \\
		0.14725 -0.20625 \\
		0.14341 -0.056026 \\
		0.14653 0.091696 \\
		0.15704 0.24309 \\
		0.1766 0.40443 \\
		0.20958 0.58451 \\
		0.26325 0.79505 \\
		0.35327 1.0523 \\
		0.51507 1.3871 \\
		0.83872 1.8413 \\
		1.5789 2.4391 \\
		3.4584 2.7207 \\
		5.6487 -0.029952 \\
		3.4032 -2.6598 \\
		1.6001 -2.3742 \\
		0.87461 -1.8007 \\
		0.55243 -1.3626 \\
		0.38863 -1.0367 \\
		0.29682 -0.78361 \\
		0.24231 -0.57652 \\
		0.20948 -0.39849 \\
		0.19078 -0.23837 \\
		0.18248 -0.088168 \\
		0.1831 0.058459 \\
		0.1927 0.20727 \\
		0.21298 0.36429 \\
		0.24783 0.53675 \\
		0.305 0.73417 \\
		0.39998 0.96997 \\
		0.5656 1.263 \\
		0.87765 1.634 \\
		1.5232 2.0623 \\
		2.8897 2.1891 \\
		4.5394 0.55621 \\
		3.5883 -1.8047 \\
		1.9616 -2.1072 \\
		1.1202 -1.7377 \\
		0.71537 -1.3589 \\
		0.50318 -1.0524 \\
		0.38232 -0.80598 \\
		0.30953 -0.60134 \\
		0.26471 -0.42439 \\
		0.23792 -0.26506 \\
		0.22414 -0.11591 \\
		0.22115 0.029103 \\
		0.2286 0.17541 \\
		0.24792 0.32856 \\
		0.28287 0.49493 \\
		0.34099 0.68257 \\
		0.43712 0.9021 \\
		0.60161 1.1666 \\
		0.90014 1.4852 \\
		1.4782 1.8202 \\
		2.5819 1.8755 \\
		3.8862 0.70378 \\
		3.4792 -1.251 \\
		2.1343 -1.8253 \\
		1.2834 -1.6274 \\
		0.83837 -1.3189 \\
		0.59541 -1.0393 \\
		0.4544 -0.80503 \\
		0.36833 -0.60657 \\
		0.31457 -0.4332 \\
		0.28165 -0.27633 \\
		0.26368 -0.12919 \\
		0.25792 0.013822 \\
		0.26375 0.15784 \\
		0.28253 0.30808 \\
		0.31801 0.47047 \\
		0.37781 0.65219 \\
		0.47846 0.86328 \\
		0.64465 1.1098 \\
		0.94345 1.3967 \\
		1.4998 1.6714 \\
		2.4873 1.6484 \\
		3.5472 0.60218 \\
		3.2399 -1.0365 \\
		2.1177 -1.621 \\
		1.327 -1.502 \\
		0.8899 -1.2397 \\
		0.64462 -0.98713 \\
		0.49971 -0.76832 \\
		0.41039 -0.57933 \\
		0.35367 -0.41342 \\
		0.31935 -0.26079 \\
		0.30091 -0.11679 \\
		0.29561 0.023754 \\
		0.30295 0.16564 \\
		0.32455 0.31381 \\
		0.36469 0.4738 \\
		0.43199 0.65209 \\
		0.543 0.85595 \\
		0.7329 1.0912 \\
		1.0612 1.3459 \\
		1.6464 1.5379 \\
		2.5717 1.3423 \\
		3.3064 0.25187 \\
		2.8547 -1.0411 \\
		1.9165 -1.4519 \\
		1.2572 -1.3446 \\
		0.87538 -1.116 \\
		0.65424 -0.89152 \\
		0.52097 -0.69192 \\
		0.43825 -0.51596 \\
		0.3869 -0.35781 \\
		0.35616 -0.21244 \\
		0.34212 -0.072983 \\
		0.34207 0.064255 \\
		0.3562 0.20364 \\
		0.387 0.34964 \\
		0.44016 0.507 \\
		0.52653 0.68063 \\
		0.66638 0.87392 \\
		0.89731 1.082 \\
		1.2851 1.2679 \\
		1.9113 1.2857 \\
		2.6603 0.81149 \\
		2.8941 -0.24155 \\
		2.3195 -1.0525 \\
		1.6138 -1.2428 \\
		1.1258 -1.1304 \\
		0.82829 -0.94143 \\
		0.64722 -0.75182 \\
		0.53533 -0.57683 \\
		0.46579 -0.41814 \\
		0.42415 -0.27211 \\
		0.40275 -0.13451 \\
		0.3979 -0.0013461 \\
		0.40866 0.13108 \\
		0.43645 0.26622 \\
		0.48542 0.40728 \\
		0.56368 0.55681 \\
		0.68524 0.71505 \\
		0.87472 0.87478 \\
		1.1673 1.0083 \\
		1.5994 1.0332 \\
		2.1258 0.78911 \\
		2.4637 0.15622 \\
		2.2874 -0.55929 \\
		1.7914 -0.93626 \\
		1.327 -0.98465 \\
		0.99809 -0.88248 \\
		0.78459 -0.73487 \\
		0.64768 -0.5822 \\
		0.561 -0.4349 \\
		0.50873 -0.29664 \\
		0.47996 -0.16571 \\
		0.4705 -0.039551 \\
		0.47963 0.084971 \\
		0.50682 0.20864 \\
		0.55632 0.33395 \\
		0.63324 0.45988 \\
		0.74774 0.58458 \\
		0.91312 0.69755 \\
		1.1466 0.77345 \\
		1.454 0.7578 \\
		1.7881 0.57757 \\
		2.0118 0.19577 \\
		1.9784 -0.26402 \\
	};

\end{smithchart}

%% file: tikzpics/BalancedAmplifier.tikz
\foreach\t in {-1,1} {
    \draw (0.5,0.3*\t) -- (1.25,0.3*\t) -- (1.25,0.5*\t) -- (3,0.5*\t) -- (3,0.3*\t) -- (3.5,0.3*\t);
}

\portLR{-0.5}{0.3}
\hybridcoupler00
\termination{-0.8}{-0.3}
\draw (-0.8,-0.3) node[anchor=east]{\scriptsize 50\,\textOmega};

\amplifier2{0.5}
\amplifier2{-0.5}

\portRL{4.5}{-0.3}
\hybridcoupler40
\termination{4.5}{0.3}
\draw (4.8,0.3) node[anchor=west]{\scriptsize 50\,\textOmega};

\draw (-0.75,0.75) node{\scriptsize\textcolor{blue}{\(A_0 \rightarrow\)}};
\draw (0.7,0.8) node{\scriptsize\textcolor{blue}{\(\frac{A_0}{\sqrt2} \rightarrow\)}};
\draw (0.7,-0.8) node{\scriptsize\textcolor{blue}{\(\frac{\I A_0}{\sqrt2} \rightarrow\)}};

\draw (1.475,0.85) node{\scriptsize\textcolor{red}{\(\leftarrow \frac{\Gamma A_0}{\sqrt2}\)}};
\draw (1.5,-0.9) node{\scriptsize\textcolor{red}{\(\leftarrow \frac{\I\Gamma A_0}{\sqrt2}\)}};
\draw (-0.75,-0.75) node{\scriptsize\textcolor{red}{\(\leftarrow \I\Gamma A_0\)}};

\draw (2.9,0.8) node{\scriptsize\textcolor{purple}{\(\frac{TA_0}{\sqrt2} \rightarrow\)}};
\draw (2.9,-0.8) node{\scriptsize\textcolor{purple}{\(\frac{\I TA_0}{\sqrt2} \rightarrow\)}};
\draw (5,-0.75) node{\scriptsize\textcolor{purple}{\(\I TA_0 \rightarrow\)}};

\draw (2,1.25) node{\scriptsize The balanced amplifier schematic};

%% file: tikzpics/DiplexedAmplifier.tikz
\portLR[Signal in]0{-0.3}
\portLR[Pump in]0{0.3}
\diplexer00

\draw (0.6,0) -- (3,0);

\amplifier{1.5}0

\diplexer30
\portRL[Signal \& idler out]{3.6}{-0.3}
\portRL[Pump out]{3.6}{0.3}

\draw (-1,0.75) node{\scriptsize\textcolor{red}{\(A\pump \rightarrow\)}};
\draw (-1,-0.75) node{\scriptsize\textcolor{blue}{\(A\signal \rightarrow\)}};
\draw (1.05,0.25) node{\scriptsize\textcolor{red}{\(A\pump \rightarrow\)}};
\draw (1.05,-0.25) node{\scriptsize\textcolor{blue}{\(A\signal \rightarrow\)}};
\draw (2.5,-0.4) node{\scriptsize\textcolor{red}{\(A\pump \rightarrow\)}};
\draw (2.4,-0.7) node{\scriptsize\textcolor{blue}{\(\sqrt{G+1}A\signal \rightarrow\)}};
\draw (2.4,-1) node{\scriptsize\textcolor{purple}{\(\sqrt{G}A\idler \rightarrow\)}};
\draw (4.5,0.75) node{\scriptsize\textcolor{red}{\(A\pump \rightarrow\)}};
\draw (4.5,-0.7) node{\scriptsize\textcolor{blue}{\(\sqrt{G+1}A\signal \rightarrow\)}};
\draw (4.5,-1) node{\scriptsize\textcolor{purple}{\(\sqrt{G}A\idler \rightarrow\)}};

\draw (1.8,1) node{\scriptsize The diplexed TWPA schematic};

%% file: tikzpics/LayoutGen4a.tikz
\draw (-1.5,0.3) -- (-0.5,0.3);
\draw (3,-0.3) -- (3.5,-0.3);
\foreach\t in {-1,1} {
    \draw (0.5,0.3*\t) -- (0.75,0.3*\t) -- (0.75,0.5*\t) -- (1.75,0.5*\t) -- (1.75,0.3*\t) -- (2,0.3*\t);
}

\portLR[]{-1.5}0
\draw (-2,-0.4) node{\scriptsize Signal in};
\portLR[]{-1.5}{0.6}
\draw (-2,1) node{\scriptsize Pump in};
\diplexer{-1.5}{0.3}
\portRL[]{4.1}{-0.6}
\draw (4.6,-1.1) node{\scriptsize Signal \& idler out};
\portRL[]{4.1}{0}
\draw (4.6,0.4) node{\scriptsize Pump out};
\diplexer{3.5}{-0.3}

\hybridcoupler00
\termination{-0.8}{-0.3}
\draw (-0.7,-0.45) node[anchor=north]{\scriptsize 50\,\textOmega};

\amplifier1{0.5}
\amplifier1{-0.5}

\hybridcoupler{2.5}0
\termination{3}{0.3}
\draw (3.1,0.45) node[anchor=south]{\scriptsize 50\,\textOmega};

\draw (1.3,1.25) node{\scriptsize The diplexed \& balanced TWPA schematic};

%% file: tikzpics/LayoutGen4b.tikz

\diplexedamplifier21
\diplexedamplifierUD2{-1}

\portLR[]{-0.5}{0.3}
\draw (-0.8,0.7) node{\scriptsize Signal in};
\hybridcoupler00
\termination{-0.8}{-0.3}
\draw (-0.7,-0.45) node[anchor=north]{\scriptsize 50\,\textOmega};

\portRL[]{5}{-0.3}
\draw (5.3,-0.7) node{\scriptsize Signal out};
\hybridcoupler{4.5}0
\termination{5}{0.3}
\draw (5.1,0.45) node[anchor=south]{\scriptsize 50\,\textOmega};

\foreach\t in {-1,1} {
    \draw (0.5,0.3*\t) -- (0.85,0.3*\t) -- (0.85,0.7*\t) -- (1.25,0.7*\t);
    \draw (3.25,0.7*\t) -- (3.6,0.7*\t) -- (3.6,0.3*\t) -- (4,0.3*\t);
}

\portLR[Pump 1 in]{1.25}{1.3}
\portLR[Pump 2 in]{1.25}{-1.3}
\portRL[Pump 1 out]{3.25}{1.3}
\portRL[Pump 2 out]{3.25}{-1.3}

\draw (2.3,2) node{\scriptsize The single layered WIF-TWPA};

%% file: tikzpics/Design7d_IdlerReflectionCoefficient.tikz
\definecolor{mycolor1}{rgb}{0.00000,0.44700,0.74100}%

\begin{smithchart}[
    xticklabels={{0.2},{},{1},{},{5}},
    title = {\massatext[0.7]{Diplexed \& balanced TWPA idler reflection coefficient for \(I\pump = 100\)\,nA}}
]

	\addplot[color=mycolor1, line width=1, forget plot] table[row sep=crcr] {
		0.93626 -0.55404 \\
		1.1437 -0.60427 \\
		1.4051 -0.56746 \\
		1.6718 -0.38414 \\
		1.8251 -0.041573 \\
		1.7529 0.3465 \\
		1.5032 0.60849 \\
		1.2059 0.69755 \\
		0.9541 0.66363 \\
		0.76875 0.56798 \\
		0.64086 0.44752 \\
		0.55658 0.31928 \\
		0.50504 0.18954 \\
		0.47947 0.059594 \\
		0.47674 -0.071559 \\
		0.49708 -0.20583 \\
		0.54441 -0.34521 \\
		0.62757 -0.49056 \\
		0.76278 -0.63857 \\
		0.97662 -0.77356 \\
		1.303 -0.84674 \\
		1.7433 -0.74101 \\
		2.137 -0.30299 \\
		2.1413 0.36145 \\
		1.7347 0.81738 \\
		1.2661 0.92481 \\
		0.91905 0.83426 \\
		0.69459 0.6813 \\
		0.5536 0.5188 \\
		0.4686 0.35766 \\
		0.41933 0.20541 \\
		0.39693 0.058283 \\
		0.39814 -0.089907 \\
		0.42111 -0.236 \\
		0.46997 -0.39094 \\
		0.55726 -0.55434 \\
		0.70101 -0.72483 \\
		0.93407 -0.88719 \\
		1.3015 -0.98616 \\
		1.8115 -0.87754 \\
		2.2612 -0.37041 \\
		2.2349 0.37358 \\
		1.7728 0.83644 \\
		1.2864 0.922 \\
		0.94097 0.82296 \\
		0.7209 0.66813 \\
		0.58408 0.5058 \\
		0.50067 0.3497 \\
		0.45329 0.20135 \\
		0.43159 0.060087 \\
		0.43492 -0.081255 \\
		0.46006 -0.2207 \\
		0.5106 -0.36388 \\
		0.59799 -0.50621 \\
		0.73551 -0.64712 \\
		0.93485 -0.76906 \\
		1.2305 -0.81557 \\
		1.5724 -0.72631 \\
		1.8758 -0.4348 \\
		1.9825 0.046699 \\
		1.7615 0.43579 \\
		1.4803 0.59728 \\
		1.1703 0.63254 \\
		0.95579 0.58541 \\
		0.80235 0.4871 \\
		0.69835 0.37583 \\
		0.63032 0.26154 \\
		0.591 0.14881 \\
		0.57446 0.037633 \\
		0.57838 -0.071556 \\
		0.60252 -0.17874 \\
		0.64859 -0.28313 \\
		0.72053 -0.38182 \\
		0.82189 -0.46998 \\
		0.95779 -0.53656 \\
		1.1289 -0.56355 \\
		1.3243 -0.52598 \\
		1.5115 -0.40109 \\
		1.6372 -0.19048 \\
		1.6535 0.060982 \\
		1.5575 0.28349 \\
		1.392 0.42868 \\
		1.209 0.49002 \\
		1.0422 0.48627 \\
		0.90439 0.44091 \\
		0.79756 0.37002 \\
		0.7185 0.28537 \\
		0.66335 0.1931 \\
		0.62902 0.096307 \\
		0.6137 -0.0036691 \\
		0.61711 -0.1064 \\
		0.64064 -0.21169 \\
		0.68774 -0.3189 \\
		0.76443 -0.42574 \\
		0.87978 -0.52605 \\
		1.0453 -0.60542 \\
		1.2695 -0.63395 \\
		1.5407 -0.56044 \\
		1.7925 -0.33037 \\
		1.8987 0.041011 \\
		1.7799 0.41441 \\
		1.5094 0.64436 \\
		1.2185 0.70987 \\
		0.97894 0.66776 \\
		0.80289 0.57299 \\
		0.68025 0.45724 \\
		0.59803 0.33507 \\
		0.54627 0.21197 \\
		0.51859 0.089129 \\
		0.51179 -0.034227 \\
		0.52544 -0.15968 \\
		0.56196 -0.28898 \\
		0.62739 -0.42326 \\
		0.73284 -0.5613 \\
		0.89691 -0.69378 \\
		1.145 -0.79785 \\
		1.4995 -0.80756 \\
		1.9218 -0.60231 \\
		2.1969 -0.09496 \\
		2.0755 0.50118 \\
		1.6655 0.84644 \\
		1.2503 0.90742 \\
		0.9447 0.82035 \\
		0.74191 0.68215 \\
		0.61144 0.53405 \\
		0.52915 0.38958 \\
		0.47993 0.25142 \\
		0.45555 0.11848 \\
		0.45016 -0.011847 \\
		0.46426 -0.14221 \\
		0.49937 -0.27505 \\
		0.55999 -0.4138 \\
		0.65726 -0.55764 \\
		0.80815 -0.70277 \\
		1.0399 -0.83122 \\
		1.3844 -0.8904 \\
		1.8327 -0.75998 \\
		2.2211 -0.304 \\
		2.225 0.3639 \\
		1.8295 0.83275 \\
		1.3617 0.9617 \\
		1.0052 0.8928 \\
		0.76707 0.75349 \\
		0.61581 0.59829 \\
		0.52021 0.44692 \\
		0.46156 0.30333 \\
		0.42902 0.16647 \\
		0.41687 0.033716 \\
		0.42289 -0.09799 \\
		0.4477 -0.23177 \\
		0.49495 -0.37057 \\
		0.57237 -0.51664 \\
		0.69393 -0.66976 \\
		0.88334 -0.82207 \\
		1.1766 -0.94378 \\
		1.6049 -0.94927 \\
		2.1155 -0.66975 \\
		2.3999 -0.010468 \\
		2.1573 0.68299 \\
		1.6321 1.0029 \\
		1.1714 1.0053 \\
		0.85761 0.87493 \\
		0.65765 0.71026 \\
		0.53137 0.54575 \\
		0.45207 0.39007 \\
		0.40411 0.24311 \\
		0.37886 0.10202 \\
		0.37213 -0.036847 \\
		0.38228 -0.17798 \\
		0.41273 -0.32463 \\
		0.46855 -0.48174 \\
		0.56143 -0.65355 \\
		0.71359 -0.84171 \\
		0.96609 -1.0361 \\
		1.3872 -1.1837 \\
		2.0333 -1.1052 \\
		2.6826 -0.46622 \\
		2.6426 0.5763 \\
		1.9601 1.1661 \\
		1.3173 1.2084 \\
		0.90584 1.0424 \\
		0.661 0.83818 \\
		0.51382 0.64378 \\
		0.42376 0.46702 \\
		0.36907 0.30514 \\
		0.33831 0.15308 \\
		0.32584 0.0055989 \\
		0.32967 -0.14255 \\
		0.35093 -0.29686 \\
		0.39431 -0.46358 \\
		0.4702 -0.65002 \\
		0.5997 -0.864 \\
		0.82616 -1.1088 \\
		1.2381 -1.3574 \\
		1.9798 -1.4453 \\
		2.9681 -0.84147 \\
		3.1058 0.60361 \\
		2.1643 1.41 \\
		1.348 1.4 \\
		0.8841 1.1619 \\
		0.62966 0.91251 \\
		0.48441 0.69247 \\
		0.39852 0.50105 \\
		0.34772 0.33056 \\
		0.31999 0.17336 \\
		0.30957 0.022799 \\
		0.3145 -0.12713 \\
		0.33584 -0.28239 \\
		0.37811 -0.44958 \\
		0.45124 -0.63639 \\
		0.57555 -0.85147 \\
		0.79305 -1.1002 \\
		1.1915 -1.3616 \\
		1.9247 -1.4859 \\
		2.9595 -0.93529 \\
		3.1803 0.55279 \\
		2.2187 1.4246 \\
		1.3712 1.4207 \\
		0.89607 1.176 \\
		0.63921 0.92191 \\
		0.49452 0.6996 \\
		0.41028 0.50759 \\
		0.36167 0.33771 \\
		0.33652 0.18213 \\
		0.32914 0.034306 \\
		0.33766 -0.11144 \\
		0.36323 -0.26042 \\
		0.41035 -0.41792 \\
		0.48851 -0.58931 \\
		0.616 -0.77884 \\
		0.82756 -0.98402 \\
		1.1862 -1.1744 \\
		1.7755 -1.2234 \\
		2.5231 -0.80195 \\
		2.7907 0.23545 \\
		2.2135 1.0509 \\
		1.5038 1.2202 \\
		1.0279 1.0875 \\
		0.7466 0.88739 \\
		0.5807 0.69135 \\
		0.48151 0.51356 \\
		0.42301 0.35278 \\
		0.39101 0.20426 \\
		0.38014 0.063739 \\
		0.38704 -0.073255 \\
		0.41214 -0.21054 \\
		0.45892 -0.35145 \\
		0.53472 -0.49801 \\
		0.65148 -0.6489 \\
		0.83437 -0.79971 \\
		1.1114 -0.91871 \\
		1.5059 -0.92861 \\
		1.972 -0.69355 \\
		2.2557 -0.1335 \\
		2.1052 0.49046 \\
		1.6766 0.833 \\
		1.2592 0.89001 \\
		0.95496 0.80466 \\
		0.75253 0.67058 \\
		0.62129 0.52687 \\
		0.53777 0.38652 \\
		0.48652 0.25212 \\
		0.46064 0.12333 \\
		0.45496 -0.0018694 \\
		0.46822 -0.12552 \\
		0.50172 -0.24929 \\
		0.55943 -0.37392 \\
		0.64855 -0.498 \\
		0.78006 -0.61545 \\
		0.96803 -0.71022 \\
		1.223 -0.74743 \\
		1.5298 -0.66714 \\
		1.8073 -0.41047 \\
		1.9137 -0.0088139 \\
		1.7816 0.37562 \\
		1.5069 0.60267 \\
		1.2217 0.66722 \\
		0.9893 0.62991 \\
		0.81828 0.54315 \\
		0.69821 0.43638 \\
		0.61688 0.32328 \\
		0.56562 0.20957 \\
		0.53691 0.09663 \\
		0.52826 -0.015469 \\
		0.53823 -0.12772 \\
		0.56889 -0.23975 \\
		0.62303 -0.35149 \\
		0.70642 -0.46049 \\
		0.8273 -0.56002 \\
		0.99494 -0.63517 \\
		1.2136 -0.65745 \\
		1.4663 -0.58351 \\
		1.692 -0.37564 \\
		1.7927 -0.05618 \\
		1.7134 0.26735 \\
		1.5053 0.48619 \\
		1.2639 0.57476 \\
		1.0507 0.56729 \\
		0.88423 0.50518 \\
		0.7621 0.41643 \\
		0.67619 0.31614 \\
		0.61895 0.21149 \\
		0.58495 0.10535 \\
		0.57101 -0.001448 \\
		0.57602 -0.10902 \\
		0.60083 -0.21761 \\
		0.64845 -0.3269 \\
		0.72436 -0.4349 \\
		0.83676 -0.53588 \\
		0.99574 -0.61654 \\
		1.2083 -0.65005 \\
		1.4633 -0.59214 \\
		1.7048 -0.39674 \\
		1.8282 -0.072917 \\
		1.76 0.27105 \\
		1.5451 0.50888 \\
		1.2903 0.60379 \\
		1.0661 0.59546 \\
		0.89301 0.52909 \\
		0.76708 0.43593 \\
		0.68074 0.33125 \\
		0.62359 0.22329 \\
		0.59029 0.11449 \\
		0.57743 0.0054072 \\
		0.58333 -0.10465 \\
		0.61005 -0.21562 \\
		0.66043 -0.32764 \\
		0.74072 -0.43879 \\
		0.86045 -0.54299 \\
		1.0317 -0.62524 \\
		1.2632 -0.65395 \\
		1.5407 -0.57543 \\
		1.7897 -0.33614 \\
		1.8845 0.036577 \\
		1.7572 0.3965 \\
		1.4932 0.60922 \\
		1.2183 0.66608 \\
		0.99514 0.62458 \\
		0.83233 0.53579 \\
		0.71962 0.42849 \\
		0.64494 0.3159 \\
		0.5991 0.20321 \\
		0.57628 0.091849 \\
		0.57346 -0.01831 \\
		0.59001 -0.12789 \\
		0.62755 -0.23725 \\
		0.69015 -0.34563 \\
		0.78463 -0.44968 \\
		0.92057 -0.54047 \\
		1.1079 -0.59824 \\
		1.3471 -0.58573 \\
		1.6051 -0.45093 \\
		1.7902 -0.16767 \\
		1.7943 0.19203 \\
		1.6128 0.48191 \\
		1.352 0.62209 \\
		1.1086 0.63563 \\
		0.91902 0.5764 \\
		0.78308 0.48389 \\
		0.69023 0.37911 \\
		0.63019 0.27153 \\
		0.59552 0.16476 \\
		0.58165 0.059732 \\
		0.58644 -0.04365 \\
		0.6098 -0.14566 \\
		0.65359 -0.24603 \\
		0.72169 -0.34304 \\
		0.82 -0.43201 \\
		0.95558 -0.50257 \\
		1.1331 -0.53478 \\
		1.345 -0.49647 \\
		1.5536 -0.35215 \\
		1.684 -0.098117 \\
		1.6671 0.19858 \\
		1.5112 0.434 \\
		1.2932 0.55423 \\
		1.0833 0.57304 \\
		0.91255 0.52804 \\
		0.78461 0.44989 \\
		0.69343 0.35674 \\
		0.63162 0.25768 \\
		0.59338 0.15668 \\
		0.575 0.055114 \\
		0.57476 -0.046883 \\
		0.59285 -0.14953 \\
		0.63137 -0.25274 \\
		0.69467 -0.35527 \\
		0.78962 -0.45313 \\
		0.92541 -0.53664 \\
		1.1107 -0.58537 \\
		1.3436 -0.56237 \\
		1.5886 -0.41818 \\
		1.7552 -0.13587 \\
		1.745 0.2129 \\
		1.5593 0.48945 \\
		1.2996 0.62111 \\
		1.0569 0.63018 \\
		0.86595 0.56755 \\
		0.72724 0.47098 \\
		0.63088 0.36077 \\
	};

\end{smithchart}

%% file: tikzpics/LayoutGen5a.tikz

\foreach\yoff in {-1.25,1.25} {
    \diplexedamplifier0{\yoff+0.625}
    \diplexedamplifierUD0{\yoff-0.625}
    
    \hybridcoupler{-1.5}{\yoff}
    \termination{-2.3}{\yoff-0.3}
    
    \hybridcoupler{2}{\yoff}
    \termination{2.5}{\yoff+0.3}
    
    \foreach\t in {-1,1} {
        \foreach\x in {-1,1.25} {
            \draw (\x,\yoff+0.3*\t) -- (\x+0.25,\yoff+0.3*\t);
        }
    }
}

\hybridcoupler{-3}{0}
\termination{-3.8}{-0.3}
\hybridcoupler{3.5}{0}
\termination{4}{-0.3}

\draw[blue,dashed] (-3.65,-0.6) -- (-3.65,-1.2) node[anchor=north]{\scriptsize1};
\draw[blue,dashed] (-2.15,0.6) -- (-2.15,0.2) (-2.15,0) node{\scriptsize2} (-2.15,-0.2) -- (-2.15,-1.3);
\draw[blue,dashed] (4.15,-0.6) -- (4.15,-1.2) node[anchor=north]{\scriptsize3};

\draw (-2.5,0.3) -- (-2.4,0.3) -- (-2.4,1.55) -- (-2,1.55);
\draw (-2.5,-0.3) -- (-2.4,-0.3) -- (-2.4,-0.95) -- (-2,-0.95);
\draw (2.5,0.95) -- (2.9,0.95) -- (2.9,0.3) -- (3,0.3);
\draw (2.5,-1.55) -- (2.9,-1.55) -- (2.9,-0.3) -- (3,-0.3);

\portLR[P1 in]{-0.75}{2.2}
\portLR[P2 in]{-0.75}{0.35}
\portLR[P3 in]{-0.75}{-0.3}
\portLR[P4 in]{-0.75}{-2.15}
\portRL[P1 out]{1.25}{2.2}
\portRL[P2 out]{1.25}{0.35}
\portRL[P3 out]{1.25}{-0.3}
\portRL[P4 out]{1.25}{-2.15}

\portUD[]{-3.6}{0.3}
\draw (-3.8,0.8) node[anchor=south west]{\scriptsize Signal in};
\portUD[]{4.1}{0.3}
\draw (4.3,0.8) node[anchor=south east]{\scriptsize Idler out};
\foreach\x in {-3.6,4} {
    \draw (\x,0.3) -- (\x+0.1,0.3);
}

%% file: tikzpics/PumpPhaseGenerating.tikz
\portLR[Pump in/out]0{0.6}

\hybridcoupler[180]{0.5}{0.3}
\termination{-0.3}0
\hybridcoupler[180]3{0.9}
\termination{2.2}{1.2}
\hybridcoupler[180]3{-0.3}
\termination{2.2}0

\draw (1,0.6) -- (2.5,0.6);
\draw (1,0) -- (1.1,0) -- (1.1,-0.6) -- (2.5,-0.6);

\foreach\t in {1,2,3,4} {
    \portRL[Pump \t\ in/out]{3.5}{1.8-0.6*\t}
}

\draw (-0.5,1) node{\scriptsize\textcolor{blue}{\(A\pump \rightarrow\)}};
\draw (1.75,0.85) node{\scriptsize\textcolor{red}{\(A\pump/\sqrt2 \rightarrow\)}};
\draw (1.75,-0.4) node{\scriptsize\textcolor{red}{\(A\pump/\sqrt2 \rightarrow\)}};
\draw (4.25,1.5) node{\scriptsize\textcolor{purple}{\(-A\pump/2 \rightarrow\)}};
\draw (4.25,0.9) node{\scriptsize\textcolor{purple}{\(A\pump/2 \rightarrow\)}};
\draw (4.25,0.3) node{\scriptsize\textcolor{purple}{\(-A\pump/2 \rightarrow\)}};
\draw (4.25,-0.3) node{\scriptsize\textcolor{purple}{\(A\pump/2 \rightarrow\)}};